\documentclass[12pt]{article}

\usepackage{preamble}

\title{Latent Impact and Differential Item Functioning Analysis for Asymmetric IRT Models}

\author[1]{Gabriel Wallin}
\author[2]{ Qi Huang}

\affil[1]{School of Mathematical Sciences, Lancaster University}
\affil[2]{Department of Educational Studies, Purdue University}
\date{}

\begin{document}


\maketitle


\begin{abstract}
Differential item functioning (DIF) arises alongside latent population heterogeneity in many applications, and both must be accounted for when assessing measurement invariance. In many practical settings, however, the comparison groups are unobserved and anchor items are unknown. A further challenge is that item response theory models traditionally assume symmetric link functions, yet empirical response processes may exhibit substantial asymmetry. This paper proposes a general framework for jointly analysing impact and DIF under asymmetric item response models. Unobserved group differences are represented by latent classes within a mixture item response model, while item-specific shifts capture DIF effects. Assuming the number of DIF items is relatively small, an $\ell_1$-regularised estimator is used to simultaneously identify the latent classes and select DIF items without requiring observed group labels or pre-specified anchor items. A simulation study evaluates recovery of impact, item parameters, and DIF effects across a range of configurations. The method is illustrated using two empirical applications from educational testing. In one dataset, the selected model reveals both impact and item-level DIF, whereas in the other, the results indicate substantial impact but little evidence of item-level DIF.

\end{abstract}

\newpage

\section{Introduction}\label{sec:intro}

Item response theory (IRT) is a popular psychometric framework for
measuring latent psychological and educational constructs from binary or
polytomous item response data \citep{embretson2013item}. Common for traditional IRT is that the item characteristic curve (ICC) is symmetric:
conditional on the latent trait, the probability of a correct response follows a logistic or probit function whose point of maximum slope lies at the inflection point. This assumption is mathematically convenient but often conflicts with empirical response processes. \citet{samejima2000logistic} introduced the logistic positive-exponent family, demonstrating that asymmetric ICCs frequently better capture human behaviour. Recent work on asymmetric IRT models have strengthened this finding \citep{bolt2022item, lee2018alternative, lee2018asymmetric}.

A stream of asymmetric IRT alternatives has since emerged. Early examples include the flexible ICCs of \citet{bazan2006skew} and Bayesian extensions by \citet{bolfarine2010bayesian}. \citet{molenaar2015heteroscedastic} proposed the heteroscedastic latent-trait model, and \citet{martin2006irt} introduced an ability-based guessing model that produces skewed lower tails. More recently, \citet{shim2023a} proposed parsimonious asymmetric models with the
complementary log-log (CLL) link, and \citet{shim2024parsimonious} developed an asymmetric IRT model with a negative log-log link function. Furthermore, \citet{verkuilen2023gumbel} introduced the Gumbel-Reverse Gumbel model for binary data.

Despite this methodological progress, DIF analysis under asymmetric ICCs remains largely unexplored. \cite{feuerstahler2026defining} established asymmetry as a fundamental item-level property that affects latent trait estimation in systematic ways, yet the inferential question of how asymmetric item behavior interacts with group differences in latent distributions, and specifically how to disentangle latent impact from item-level DIF under asymmetric ICCs, has not been addressed. When a symmetric model is imposed on asymmetric data, the mismatch in item characteristic curve shape may be absorbed by the latent trait metric or by item parameter adjustments \citep{bolt2022item}, creating a risk of confounding between true measurement non-invariance and model misfit. The present paper fills this gap by proposing a unified framework for latent DIF detection applicable to asymmetric IRT models that are themselves identifiable, which is a condition that may not be satisfied by all asymmetric specifications, as \cite{gonzalez2025identifiability} formally demonstrates for the logistic positive exponent model and related variants. 

The problem of DIF analysis for asymmetric IRT models could be further complicated in practice by two sources of missing information. First, comparison groups may not be directly observable. The latent classes can represent well-defined but unrecorded subpopulations, such as country, gender, or language background, but they may equally capture more diffuse heterogeneity without a direct observable counterpart, such as speededness, shifts in response behaviour, or aberrant responding. When focal and reference
groups are latent, standard DIF methods that require observed group labels cannot be applied. Latent DIF analysis addresses this by modelling group membership as a latent class variable
\citep{cho2016ncme, de2011explanatory}. Second, anchor items may be unavailable or misspecified. Errors in anchor specification propagate bias into all subsequent inferences. Lasso-type regularisation methods have been proposed to identify anchors in the symmetric setting when groups are observed \citep{magis2015detection, tutz2015penalty, belzak2020improving, bauer2020simplifying}, and \citet{wallin2024dif} extended this to the case of missing group labels under the symmetric two-parameter logistic model.

To the best of our knowledge, no existing method addresses DIF detection under asymmetric ICCs while simultaneously treating both group membership and anchor items as unknown. The present paper fills this gap. The proposed framework is general with respect to the choice of asymmetric link function: any asymmetric IRF that is itself identifiable and for which a tractable marginal likelihood can be evaluated may in principle be embedded within the latent-class mixture and regularised estimation structure developed here. As a concrete and well-motivated instantiation, we adopt the CLL link with the Gumbel error distribution \citep{shim2023a}. The CLL model exhibits constant negative asymmetry, consistent with disjunctive response processes in which multiple pathways to a correct response reduce the steepness of the ICC in the upper tail. This might be a theoretically plausible characterisation of multiple-choice mathematics items where varied solution strategies are available. This choice also yields computational coherence: the CLL link is the canonical link function associated with the Gumbel error distribution \cite{shim2023a}, so specifying Gumbel-distributed latent classes ensures consistency between the measurement and structural components of the model. In the current paper, we thus use CLL link for our baseline asymmetric IRT model, embed it within a latent-class mixture framework, and propose an $\ell_1$-penalised marginal maximum likelihood estimator that simultaneously estimates item parameters, identifies latent
classes, and selects DIF items without requiring pre-specified anchors or observed groups. 

The transition from the symmetric two-parameter logistic model to the proposed CLL-Gumbel framework introduces several non-standard features that affect estimation and interpretation. First, the asymmetric CLL link implies a non-standard score function, and the interaction between this score function and the $\ell_1$ penalty produces a regularisation path whose geometry differs from the logistic case: the CLL score is bounded away from zero for items near the lower asymptote but decays sharply in the upper tail, which changes the relative influence of high- and low-difficulty items on anchor selection. Second, the Gumbel distribution for the latent trait is asymmetric and differs from the normal in its tail behaviour, and the class-specific Gumbel parameters interact with the CLL link in a way that alters the relationship between the structural and measurement components of the model. In particular, shifts in the class-specific latent trait distribution and shifts in the DIF parameters are not interchangeable in the same way as in the logistic-normal setting, leading to different marginal likelihood behaviour. Third, the asymmetric link leads to asymmetric posterior distributions for latent class membership, which can affect the stability of the EM algorithm. In this paper, we account for these features in the model formulation and estimation procedure. 

The rest of the paper is organised as follows. Section~\ref{sec:model} presents the proposed framework and identification analysis. Section~\ref{sec:computation} describes the proximal EM algorithm and model selection procedure. Section~\ref{sec:simulation} reports the simulation study. Section~\ref{sec:empirical} presents two empirical illustrations. Section~\ref{sec:discussion} discusses limitations and extensions. Technical details are collected in the Appendix.

\section{Proposed Framework}\label{sec:model}

\subsection{Measurement Model}\label{sec:measurement}

Consider $N$ respondents answering $J$ binary items. Let $Y_{ij} \in \{0,1\}$ denote respondent $i$'s response to item $j$, collected in the vector
$\bs{Y}_i = (Y_{i1},\ldots,Y_{iJ})^\top$. The latent ability of respondent $i$ is denoted by $\theta_i$. It is assumed that the $N$ respondents belong to one of $K+1$ unobserved latent classes, which are encoded by the discrete random variable $\xi_i \in \{0,1,\ldots,K\}$, and which may induce DIF.

Let $F:\R \to [0,1]$ be a strictly increasing, differentiable CDF representing the chosen item characteristic curve, and let $g_k(\theta;\bs{\eta}_k)$ denote the density of $\theta_i$ conditional on $\xi_i = k$, parametrised by class-specific structural parameters $\bs{\eta}_k$. Conditional on $\theta_i$ and $\xi_i = k$, the proposed item response function is
\begin{equation}\label{eq:irf-general}
  P(Y_{ij} = 1 \mid \theta_i, \xi_i = k)
  = F\!\bigl(\theta_i - d_j - \delta_{jk}\bigr),
\end{equation}
where $d_j \in \R$ is the baseline item difficulty and $\delta_{jk} \in \R$ is the DIF effect for class $k$ on item $j$, with $\delta_{j0} = 0$ for all $j$ identifying the reference class. We assume throughout that the baseline IRT model obtained by setting $\delta_{jk} = 0$ for all $j$ and $k$ is identifiable; for asymmetric specifications this is not automatic and depends on the chosen $F$ \citep{gonzalez2025identifiability}. The estimation procedure developed in Section~\ref{sec:computation} requires only the score function $f(z)/[F(z)(1-F(z))]$, where $f = F'$, together with a quadrature scheme adapted to $g_k$. Different pairings $(F, g_k)$ therefore yield different specialisations of the same framework; pairings for which $g_k$ is the canonical latent distribution for $F$ give closed-form marginal expressions and thus some computational simplifications. Evaluation of the framework under alternative pairings, such as the negative log-log link of \citet{shim2024parsimonious}, is left for future work.

In the remainder of the paper we adopt the complementary log-log (CLL) link paired with class-specific Gumbel densities, since this is the canonical pairing \citep{shim2023a} and matches the disjunctive response process motivation of our empirical applications. We thus focus our presentation and analysis on the specific member of the model in \eqref{eq:irf-general} given by 
\begin{equation}\label{eq:irf}
  P(Y_{ij} = 1 \mid \theta_i, \xi_i = k)
  = \Pjk(\theta_i)
  = 1 - \exp\!\bigl[-\exp\!\bigl(\theta_i - d_j - \delta_{jk}\bigr)\bigr].
\end{equation}
For $k \geq 1$, a non-zero $\delta_{jk}$ shifts the item difficulty in the CLL metric, inducing DIF.

The CLL link $F(x) = 1 - e^{-e^x}$ is asymmetric: the inflection point lies above $0.5$ on the probability scale, and the lower tail of the ICC approaches zero more gradually than the upper tail approaches one. The score function of the CLL link is
\begin{equation}\label{eq:score}
  \frac{\partial \log \Pjk}{\partial \theta_i}
  = \frac{e^{z_{jk} - e^{z_{jk}}}}{\Pjk(1 - \Pjk)},
  \quad z_{jk} = \theta_i - d_j - \delta_{jk},
\end{equation}
which exhibits asymmetric behaviour across the latent trait range, decaying rapidly for large positive values of $z_{jk}$. This asymmetry affects the contribution of different items to the gradient underlying the regularised estimation procedure. In particular, items that are very easy for the reference group contribute less gradient signal in regions where the response probability is close to one. As a consequence, the interaction between the CLL score function and the $\ell_1$ penalty may lead to a selection pattern for DIF effects that differs from the symmetric logistic case. We investigate this behaviour in the simulation study of Section~\ref{sec:simulation}.

The framework developed in this paper is general with respect to the choice of asymmetric link function. The score function entering the proximal EM gradient (derived in Section 3 and Appendix A) is the only component of the estimation algorithm where the link function appears explicitly; substituting an alternative asymmetric link therefore requires only replacing this score function and adopting a compatible latent trait distribution.

\subsection{Structural Model}\label{sec:structural}

The latent classes follow a categorical distribution,
\[
  \xi_i \sim \mathrm{Categorical}\!\bigl(\{0,\ldots,K\},\;(\nu_0,\ldots,\nu_K)\bigr),
  \quad \sum_{k=0}^{K} \nu_k = 1,\quad \nu_k \geq 0.
\]
Conditional on $\xi_i = k$, the latent trait follows a Gumbel distribution
with class-specific location $\mu_k$ and scale $\sigma_k$:
\begin{equation}\label{eq:gumbel}
  \theta_i \mid \xi_i = k \sim \Gumbel(\mu_k, \sigma_k),
  \quad
  g(\theta;\mu_k,\sigma_k)
  = \frac{1}{\sigma_k}
    \exp\!\Bigl\{-z_k - e^{-z_k}\Bigr\},
  \quad z_k = \frac{\theta - \mu_k}{\sigma_k}.
\end{equation}
To fix the metric, we set $\mu_0 = 0$ and $\sigma_0 = 1$ for the reference class.

When the latent trait is Gumbel-distributed and the item response follows a CLL link, the marginal item response probability has a closed-form extreme-value convolution \citep{shim2023a}. However, as in standard normal–logistic models, the marginal likelihood remains invariant under joint shifts of the latent location and item difficulty parameters. Consequently, location constraints are still required for identification. We formalise these identification considerations in Section~\ref{sec:mml}.

\subsection{Marginal Likelihood and Identifiability}\label{sec:mml}

Let $\bs{\Omega}$ denote all free parameters: the item difficulties $d_j$, DIF
parameters $\delta_{jk}$, class proportions $\nu_k$, and class-specific Gumbel
location and scale parameters $\mu_k$ and $\sigma_k$, for $j = 1,\ldots,J$ and
$k = 0,\ldots,K$. Marginalising over both $(\theta_i, \xi_i)$ yields the marginal likelihood
\begin{equation}\label{eq:marglik}
  L(\bs{\Omega})
  = \prod_{i=1}^{N}
    \sum_{k=0}^{K} \nu_k
    \int
    \prod_{j=1}^{J}
    \Pjk(\theta)^{Y_{ij}}
    \bigl[1 - \Pjk(\theta)\bigr]^{1-Y_{ij}}
    g(\theta;\mu_k,\sigma_k)\,\mathrm{d}\theta.
\end{equation}
The integral in~\eqref{eq:marglik} is approximated by numerical quadrature. We
use a fixed grid of $G = 61$ points spanning $(-8, 8)$ with Gumbel-adapted weights
$\omega_q \propto \exp\{-\rho_q - e^{-\xi_q}\}$, normalised to sum to one, where $\rho_q$ are evenly spaced nodes on the grid.

The model raises two formal identification issues, together with a broader interpretational concern about the meaning of the latent classes:

\textit{Anchor indeterminacy.} The marginal likelihood~\eqref{eq:marglik} is invariant under the joint transformation $\mu_k \mapsto \mu_k + c_k$ and $\delta_{jk} \mapsto \delta_{jk} + c_k$ for any constants $c_1,\ldots,c_K \in \R$ and all $j$, $k \geq 1$, with the reference-class parameters $\mu_0$ and $\delta_{j0}$ held fixed. This follows from the substitution $u = \theta - c_k$ in the class-$k$ integrand, which absorbs the shift in $g_k$ whenever $g_k$ is a location-scale family with location $\mu_k$. The argument is the one used by \citet{wallin2024dif} for the symmetric two-parameter logistic model with a normal trait, and it extends without modification to the present asymmetric framework because it depends only on the additive parametrisation $F(\theta - d_j - \delta_{jk})$ and on the location-scale structure of $g_k$, not on the specific form of $F$ or the baseline density $h$. Identification therefore requires additional constraints. As in \citet{wallin2024dif} we proceed under the assumption of sparsity in $\bs{\delta}$: items with $\delta_{jk} = 0$ act as within-class anchors that pin each $c_k$ to zero, provided at least one such item exists per class. This strategy makes interpretation conditional on the sparsity assumption being approximately correct. If DIF is in fact widespread, the regularised estimator may still return a stable solution, but the zero constraints induced by the penalty may effectively define the scale through an incorrect set of anchor items, in which case part of the class difference in ability may be misattributed to DIF or vice versa. This is analogous to anchor misspecification in manifest DIF analysis. 

\textit{Label switching.} The reference class is pinned by the constraints
$\delta_{j0} = 0$, $\mu_0 = 0$, $\sigma_0 = 1$. The remaining $K$ focal classes are exchangeable. Label switching does not affect the value of the maximised likelihood, but it does
complicate the comparison of solutions across random starts, since equivalent
solutions may differ only by a permutation of the class labels.
We handle this by ordering classes by proportion after convergence
($\nu_1 \geq \nu_2 \geq \ldots \geq \nu_K$).

\textit{Distributional heterogeneity versus measurement non-invariance.} An interpretational issue concerns the meaning of the latent classes themselves. In the proposed framework, a focal class with $\nu_k > 0$ and $\delta_{jk} \neq 0$ is interpreted as a subgroup of respondents for whom item $j$ functions differently. However, the same mixture structure could arise from non-Gumbel heterogeneity that the mixture is approximating, or from violations of the
CLL or single-factor assumptions. In such cases, the estimated $\delta_{jk}$ may partly reflect model misspecification rather than substantive item bias. We examine this concern in three ways. First, the simulation study includes a design without DIF (Design B), allowing us to assess whether between-class differences in the latent distribution alone can induce spurious DIF detection. Second, in the empirical illustration we use known country labels to examine the identified latent classes. Third, in Section~\ref{sec:discussion} we outline the conditions under which the latent class interpretation is substantively defensible.

\subsection{Sparsity, Regularisation, and Model Selection}\label{sec:estimation}

Under the sparsity assumption, we propose the $\ell_1$-penalised marginal maximum
likelihood estimator
\begin{equation}\label{eq:lasso}
  \widetilde{\bs{\Omega}}(\lambda)
  = \arg\min_{\bs{\Omega}}
  \Bigl\{
    -\log L(\bs{\Omega})
    + \lambda \sum_{j=1}^{J}\sum_{k=1}^{K} |\delta_{jk}|
  \Bigr\},
  \quad
  \text{s.t. } \nu_k \geq 0,\;\sum_{k=0}^K \nu_k = 1.
\end{equation}
The $\ell_1$ penalty shrinks small DIF effects exactly to zero, converting DIF
detection into a model selection problem. Under conditions on the marginal likelihood analogous to those that yield selection consistency in the linear regression setting \citep{zhao2006model}, the estimator is selection consistent; verifying these conditions for the present model is an open problem.

Following the two-stage procedure of \citet{wallin2024dif}, we select $\lambda$ by
BIC over a grid $\lambda_1 < \ldots < \lambda_M$. For each $\lambda_m$, we first
solve~\eqref{eq:lasso} to obtain $\widetilde{\bs{\Omega}}(\lambda_m)$ and the
support $\mathcal{S}_m = \{(j,k) : \tilde{\delta}_{jk}(\lambda_m) \neq 0\}$, then
refit the constrained unpenalised model
\begin{equation}\label{eq:refit}
  \widehat{\bs{\Omega}}(\lambda_m)
  = \arg\min_{\bs{\Omega}} \{-\log L(\bs{\Omega})\},
  \quad
  \text{s.t. } \delta_{jk} = 0 \text{ for all }
  (j,k) \notin \mathcal{S}_m,
\end{equation}
and compute
\[
  \BIC_{\lambda_m}
  = -2\log L\!\bigl(\widehat{\bs{\Omega}}(\lambda_m)\bigr)
    + \log(N)\,\mathrm{Card}\!\bigl(\widehat{\bs{\Omega}}(\lambda_m)\bigr).
\]
The selected tuning parameter is $\hat\lambda = \arg\min_m \BIC_{\lambda_m}$, and
$\widehat{\bs{\Omega}}(\hat\lambda)$ is the final estimator.

The confirmatory refit in~\eqref{eq:refit} serves two purposes. It removes the
shrinkage bias introduced by the $\ell_1$ penalty, so that the final parameter
estimates are asymptotically unbiased given a correctly selected support. It also
provides a log-likelihood on which the BIC is computed under a standard model
complexity count, rather than on the penalised objective whose effective degrees of
freedom are not directly interpretable. This two-stage design is analogous to the
active-set approach of \citet{wang2021using} in the manifest DIF literature, and its
theoretical justification rests on the selection consistency of the LASSO
\citep{zhao2006model} combined with the asymptotic properties of BIC for model
selection \citep{shao1997asymptotic}. An important consequence is that the bias in
the confirmatory refit is zero asymptotically provided the LASSO selects the correct
support.

\section{Computation}\label{sec:computation}

\subsection{Proximal EM Algorithm}\label{sec:proximal-em}

The optimisation problems in~\eqref{eq:lasso} and~\eqref{eq:refit} are solved by
EM algorithms \citep{dempster1977maximum, bock1981marginal}. The non-smooth $\ell_1$
penalty in~\eqref{eq:lasso} is handled by a proximal-gradient update in the M-step
\citep{parikh2014proximal}.

\textit{E-step.} At iteration $t$, compute the posterior weights
\begin{equation}\label{eq:post}
  w_{iq}^{(k,t)}
  = \frac{
      \nu_k^{(t)}
      \prod_{j=1}^{J}
      \bigl[\Pjk^{(t)}(\rho_q)\bigr]^{Y_{ij}}
      \bigl[1-\Pjk^{(t)}(\rho_q)\bigr]^{1-Y_{ij}}
      \cdot \omega_q
    }{
      \sum_{k'=0}^{K}
      \sum_{q'=1}^{G}
      \nu_{k'}^{(t)}
      \prod_{j=1}^{J}
      \bigl[\Pjk'^{(t)}(\rho_{q'})\bigr]^{Y_{ij}}
      \bigl[1-\Pjk'^{(t)}(\rho_{q'})\bigr]^{1-Y_{ij}}
      \cdot \omega_{q'}
    },
\end{equation}
where $\rho_1,\ldots,\rho_G$ are the quadrature nodes and $\omega_q$ are the
corresponding Gumbel-adapted weights.

\textit{M-step: class proportions.} The class proportion update has the closed form
\[
  \nu_k^{(t+1)}
  = \frac{1}{N}\sum_{i=1}^{N}
    \sum_{q=1}^{G} w_{iq}^{(k,t)},
  \quad k = 0, \ldots, K.
\]
This follows from the closed-form Lagrangian solution to the constrained
optimisation of the expected complete-data log-likelihood over $\nu$
\citep{wallin2024dif}, and does not require a gradient step.

\textit{M-step: item and structural parameters.} Let
$S_q^{(k)} = \sum_i w_{iq}^{(k,t)}$ and
$O_{jq}^{(k)} = \sum_i Y_{ij} w_{iq}^{(k,t)}$.
The gradient of the negative expected complete-data log-likelihood
$Q(\bs{\Omega} \mid \bs{\Omega}^{(t)})$ with respect to $d_j$ is
\begin{equation}\label{eq:grad_d}
  \frac{\partial (-Q)}{\partial d_j}
  = -\frac{1}{N}
    \sum_{k=0}^{K}\sum_{q=1}^{G}
    \bigl(O_{jq}^{(k)} - \Pjk(\rho_q) S_q^{(k)}\bigr)
    \cdot s_{jk}(\rho_q),
\end{equation}
where the CLL score is
$s_{jk}(\rho_q) = e^{z_{jk}(\rho_q) - e^{z_{jk}(\rho_q)}} / [\Pjk(\rho_q)(1-\Pjk(\rho_q))]$
with $z_{jk}(\rho_q) = \rho_q - d_j - \delta_{jk}$.
The gradient with respect to $\delta_{jk}$ has the same form
as~\eqref{eq:grad_d} restricted to class $k$; see Appendix~\ref{app:gradients} for
all expressions. Item difficulties and structural parameters are updated by
backtracking gradient descent. The backtracking is needed because the M-step
objective is not globally convex in the item parameters under the CLL link, and a
fixed step size may overshoot local curvature changes induced by the exponential
structure of the score function. Backtracking is applied within each M-step inner iteration to ensure monotone descent of the M-step objective, not across EM iterations. The EM algorithm itself guarantees non-decreasing observed log-likelihood across outer iterations by construction.

DIF parameters are updated by a proximal gradient step that combines a gradient
descent update with soft-thresholding:
\begin{equation}\label{eq:prox}
  \delta_{jk}^{(t+1)}
  = \mathcal{S}_{\alpha\lambda}\!\bigl(
      \delta_{jk}^{(t)} - \alpha \,\nabla_{\delta_{jk}}(-Q)
    \bigr),
  \quad
  \mathcal{S}_\tau(x) = \sign(x)\max(|x|-\tau,0).
\end{equation}
The step size $\alpha$ is selected by line search on the penalised
objective. This proximal update is applied element-wise, which is valid because the $\ell_1$ penalty is separable in the $\delta_{jk}$ coordinates. The negative expected complete-data log-likelihood $-Q$ is also separable across items (each item
$j$ contributes an independent term to $-Q$ given the E-step weights), so the
element-wise update is consistent with the coordinate structure.

For the confirmatory refit in~\eqref{eq:refit}, the same EM structure is used but the proximal step~\eqref{eq:prox} is replaced by a standard gradient step, and the
zero constraints from $\hat{\mathcal{S}}$ are enforced throughout by excluding the
constrained parameters from the gradient updates.

\section{Simulation Study}\label{sec:simulation}

\subsection{Design}\label{sec:sim-design}

We evaluate the finite-sample performance of the proposed framework in the two-class ($K = 1$) setting under two designs. Design A considers both impact and DIF in which the latent classes differ in both their ability distributions and the difficulty of a subset of items. Design B considers the impact structure but generates the data with no DIF items, allowing us to assess whether the procedure spuriously selects DIF items when between-class differences are confined to ability.

Each design uses $J = 25$ items and sample sizes $N \in \{500, 1000, 3000\}$ with focal class proportion $\pi \in \{0.10, 0.30, 0.50\}$, giving nine conditions per design and eighteen in total. For both designs the latent distributions are $\theta_i \mid \xi_i = 0 \sim \Gumbel(0, 1)$ and $\theta_i \mid \xi_i = 1 \sim \Gumbel(0.75, 0.80)$, and baseline item difficulties are drawn as $d_j \sim \mathrm{Uniform}(-2, 2)$. Under Design A, items 1 through 10 exhibit DIF with $\delta_j \sim \mathrm{Uniform}(0.5, 1.5)$ and items 11 through 25 are DIF-free. Under Design B all $J$ items are DIF-free. Each scenario is replicated 100 times. Design A and B are summarised in Table \ref{tab:sim_design}. 

\begin{table}[ht]
\centering
\begin{tabular}{lcccc}
\toprule
Setting & $N$ & $J$ & $\pi$ & DIF structure \\
\midrule
A: impact + DIF & $\{500,1000,3000\}$ & 25 & $\{0.1,0.3,0.5\}$ & 10 DIF items \\
B: impact only & $\{500,1000,3000\}$ & 25 & $\{0.1,0.3,0.5\}$ & 0 DIF items \\
\bottomrule
\end{tabular}
\caption{The designs considered in the simulation study.}
\label{tab:sim_design}
\end{table}

\subsection{Evaluation Criteria}\label{sec:sim-eval}

For each replication we calculate the bias and $\RMSE$ of $d_j$, $\delta_j$, $\pi$, $\mu_1$, and $\sigma_1$. For DIF item detection we calculate the true positive rate (TPR) and the false positive rate (FPR). For respondent classification we calculate the Maximum a posteriori (MAP) classification error and the area under the curve (AUC) of the Receiver Operating Characteristic (ROC) curve.

\subsection{Results -- Design A}\label{sec:sim-resultsA}

\begin{figure}[htbp]
\centering
\includegraphics[width=1\textwidth]{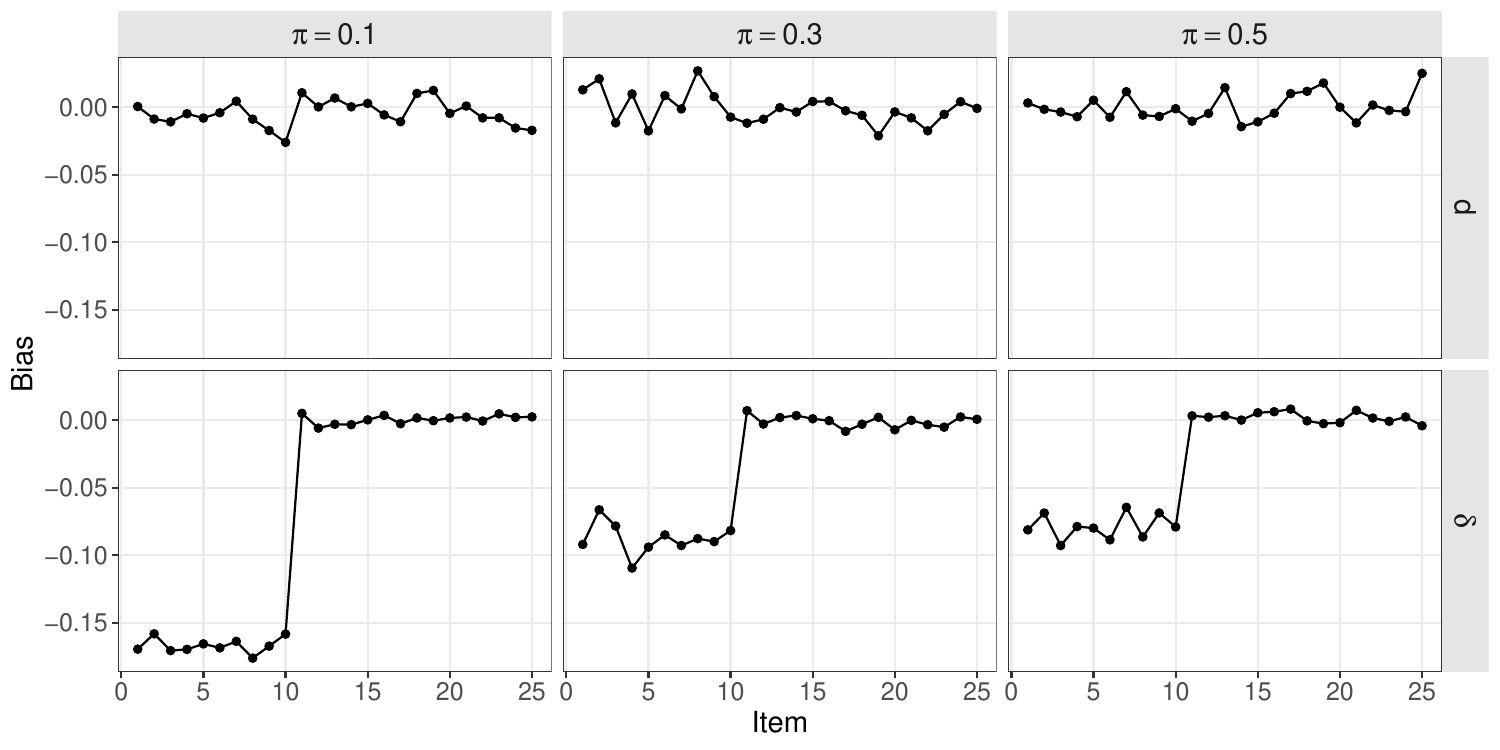}
\caption{Bias of the item parameters for $N=1000$ across values of $\pi$.}
\label{fig:bias_N1000}
\end{figure}

\begin{figure}[htbp]
\centering
\includegraphics[width=1\textwidth]{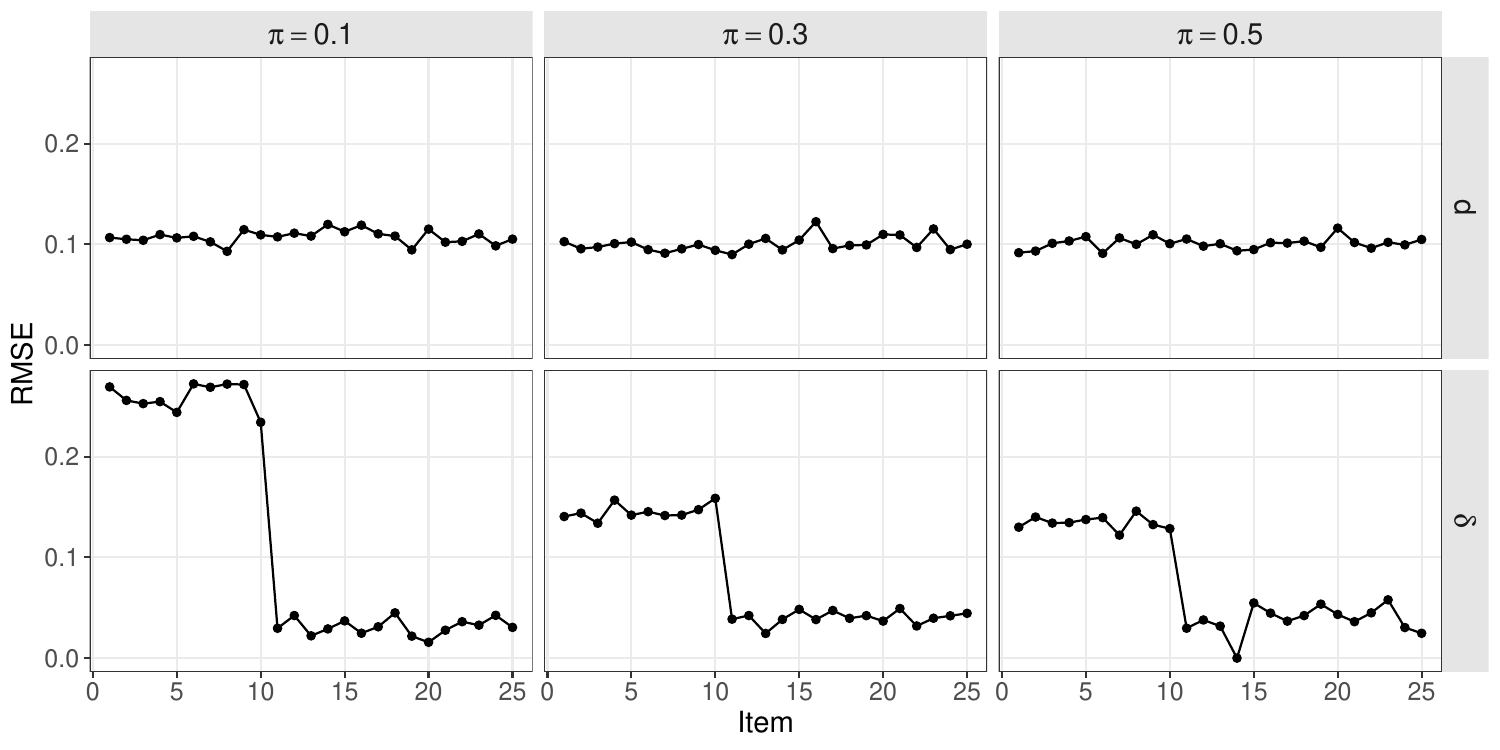}
\caption{RMSE of the item parameters for $N=1000$ across values of $\pi$.}
\label{fig:rmse_N1000}
\end{figure}

Figures \ref{fig:bias_N1000} and \ref{fig:rmse_N1000} display the bias and RMSE for $N=1000$. The bias for the $d$ parameters is close to zero across all items and values of $\pi$, with no clear systematic pattern. In contrast, the bias for the $\delta$ parameters shows a clear separation between the first ten items and the remaining items. For items 1--10, the bias is consistently negative and decreases in magnitude as $\pi$ increases, from approximately $-0.17$ at $\pi=0.1$ to around $-0.08$ at $\pi=0.5$. For items 11--25 (the DIF-free items), the bias is close to zero across all values of $\pi$.

The RMSE for the $d$ parameters is stable across items and values of $\pi$, remaining close to $0.10$. For the $\delta$ parameters, the RMSE again differs between the first ten and the remaining items. For items 1--10, the RMSE is larger and decreases as $\pi$ increases, from approximately $0.25$--$0.27$ at $\pi=0.1$ to around $0.13$--$0.15$ at $\pi=0.5$. For items 11--25, the RMSE is smaller, typically below $0.05$, and shows little variation across values of $\pi$.

These results indicate that estimation of the $\delta$ parameters is sensitive to the class proportion $\pi$, particularly for the items with nonzero effects, whereas the $d$ parameters are estimated with stable accuracy across all conditions.

Table \ref{tab:struct-params-allN} reports the mean bias and RMSE for the structural parameters across sample sizes and values of $\pi$. The bias for $\pi$ is positive across all conditions and decreases with increasing sample size, while the RMSE shows a clear reduction as $N$ increases. For $\mu_1$, the bias is close to zero in all scenarios with no systematic pattern across $\pi$ or $N$. The RMSE remains relatively stable across sample sizes, with only modest variation across conditions. For $\sigma_1$, the bias is generally small but shows some variability across values of $\pi$, including negative bias for $\pi = 0.30$ and $\pi = 0.10$ in some cases. The RMSE for $\sigma_1$ is relatively stable across all sample sizes and conditions, with no clear decreasing trend as $N$ increases. Overall, the results indicate that estimation of $\pi$ benefits from larger sample sizes, while the estimation accuracy for $\mu_1$ and $\sigma_1$ is already stable for smaller samples and does not improve substantially as $N$ increases.

\begin{table}[ht]
\centering
\caption{Mean bias and root mean squared error for structural model parameters across 100 replications, base well-specified design ($J = 25$).}
\label{tab:struct-params-allN}
\begin{tabular}{llcccccc}
\toprule
& & \multicolumn{2}{c}{$\pi = 0.10$}
& \multicolumn{2}{c}{$\pi = 0.30$}
& \multicolumn{2}{c}{$\pi = 0.50$} \\
\cmidrule(lr){3-4}\cmidrule(lr){5-6}\cmidrule(lr){7-8}
$N$ & Parameter & Bias & RMSE & Bias & RMSE & Bias & RMSE \\
\midrule
\multirow{3}{*}{500}
& $\pi$      & 0.021 & 0.055 & 0.021 & 0.056 & 0.023 & 0.057 \\
& $\mu_1$    & $-$0.000 & 0.094 & 0.014 & 0.109 & 0.013 & 0.100 \\
& $\sigma_1$ & $-$0.004 & 0.100 & 0.003 & 0.093 & 0.008 & 0.103 \\
\midrule
\multirow{3}{*}{1000}
& $\pi$      & 0.023 & 0.039 & 0.009 & 0.044 & 0.016 & 0.049 \\
& $\mu_1$    & $-$0.003 & 0.084 & 0.005 & 0.092 & $-$0.006 & 0.101 \\
& $\sigma_1$ & 0.003 & 0.101 & $-$0.014 & 0.097 & 0.008 & 0.095 \\
\midrule
\multirow{3}{*}{3000}
& $\pi$      & 0.006 & 0.027 & 0.012 & 0.038 & 0.015 & 0.044 \\
& $\mu_1$    & $-$0.002 & 0.103 & 0.001 & 0.114 & 0.001 & 0.096 \\
& $\sigma_1$ & $-$0.017 & 0.097 & $-$0.017 & 0.109 & 0.022 & 0.092 \\
\bottomrule
\end{tabular}
\end{table}

Table \ref{tab:classif-allN} reports the respondent classification error and item classification performance across sample sizes and values of $\pi$. The classification error increases with $\pi$, from about $0.08$--$0.09$ when $\pi = 0.10$ to about $0.22$--$0.23$ when $\pi = 0.50$, with little variation across sample sizes.  Note though that the performance relative to the naive classifier that puts every respondent in the baseline group shows clear improvement as the proportion in the second class increases. The AUC values are stable for $N=1000$ and $N=3000$, remaining close to $0.86$ across values of $\pi$. For $N=500$, the AUC is lower when $\pi = 0.10$, but is comparable to the larger sample sizes for $\pi = 0.30$ and $\pi = 0.50$. The TPR is high in most conditions and reaches or is close to one for $\pi = 0.30$ and $\pi = 0.50$ across all sample sizes. The main exception is the condition with $N=500$ and $\pi = 0.10$, where the TPR is $0.785$. The FPR remains small for $N=500$ and $N=1000$, but is larger for $N=3000$, especially when $\pi = 0.10$.

\begin{table}[ht]
\centering
\caption{Respondent and item classification accuracy across 100 replications, base well-specified design ($J = 25$).}
\label{tab:classif-allN}
\begin{tabular}{llccc}
\toprule
$N$ & Metric & $\pi = 0.10$ & $\pi = 0.30$ & $\pi = 0.50$ \\
\midrule
\multirow{4}{*}{500}
& Classification error & 0.089 & 0.187 & 0.224 \\
& AUC                  & 0.804 & 0.863 & 0.868 \\
& TPR                  & 0.785 & 0.998 & 0.999 \\
& FPR                  & 0.012 & 0.042 & 0.060 \\
\midrule
\multirow{4}{*}{1000}
& Classification error & 0.085 & 0.180 & 0.229 \\
& AUC                  & 0.859 & 0.866 & 0.863 \\
& TPR                  & 0.967 & 0.998 & 1.000 \\
& FPR                  & 0.038 & 0.060 & 0.066 \\
\midrule
\multirow{4}{*}{3000}
& Classification error & 0.084 & 0.184 & 0.226 \\
& AUC                  & 0.865 & 0.865 & 0.866 \\
& TPR                  & 0.998 & 1.000 & 1.000 \\
& FPR                  & 0.115 & 0.100 & 0.088 \\
\bottomrule
\end{tabular}
\end{table}

\subsection{Results -- Design B}\label{sec:sim-resultsB}

\begin{figure}[htbp]
\centering
\includegraphics[width=1\textwidth]{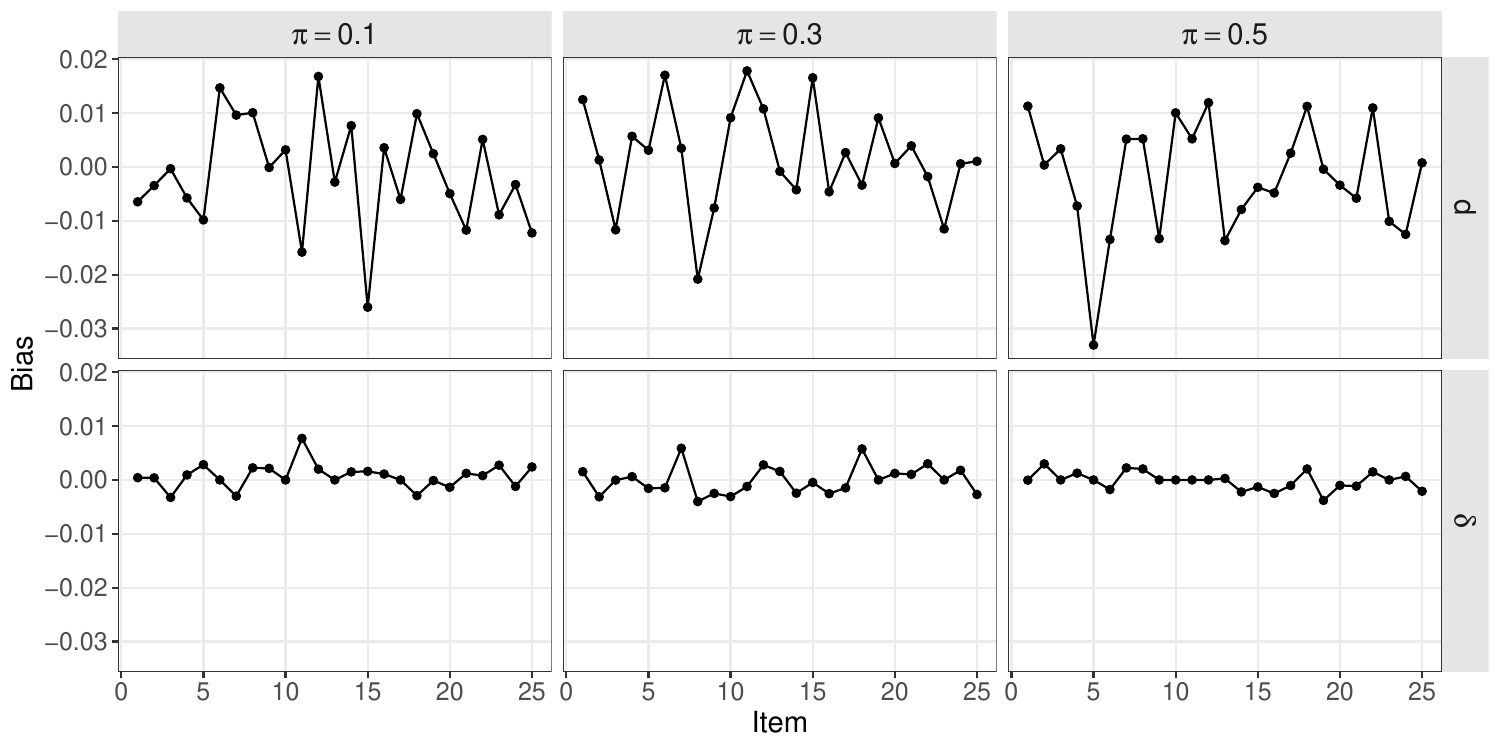}
\caption{Bias of the item parameters for $N=1000$ under Design B across values of $\pi$.}
\label{fig:bias_N1000_B}
\end{figure}

\begin{figure}[htbp]
\centering
\includegraphics[width=1\textwidth]{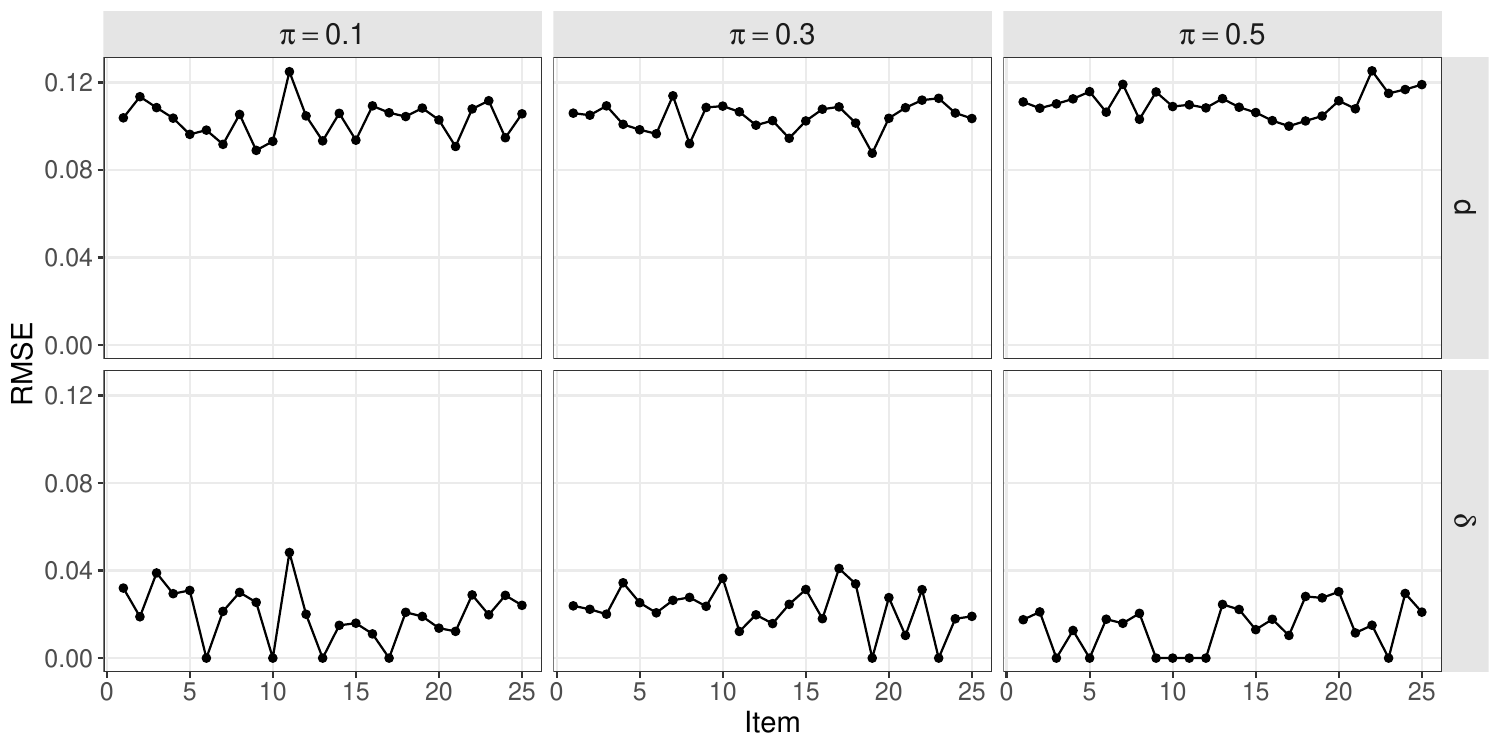}
\caption{RMSE of the item parameters for $N=1000$ under Design B across values of $\pi$.}
\label{fig:rmse_N1000_B}
\end{figure}

Figures \ref{fig:bias_N1000_B} and \ref{fig:rmse_N1000_B} display the bias and RMSE for $N=1000$ under Design B. The bias for both the $d$ and $\delta$ parameters is close to zero across all items and values of $\pi$, with no systematic pattern. In contrast to Design A, there is no separation between subsets of items, reflecting that all items are DIF-free in this design. The RMSE for the $d$ parameters is stable across items and values of $\pi$. For the $\delta$ parameters, the RMSE is small across all items, indicating that the procedure does not estimate substantial DIF effects when none are present.

Table \ref{tab:struct-params-allN-B} reports the mean bias and RMSE for the structural parameters under Design B. The bias for $\pi$ is small across all conditions and is slightly negative in most scenarios. The RMSE for $\pi$ ranges from 0.046 to 0.089 and does not show a strong monotone pattern across sample sizes. For $\mu_1$ and $\sigma_1$, the bias remains close to zero across all values of $\pi$ and $N$. Their RMSE values are also stable across conditions, mostly between 0.08 and 0.11.

\begin{table}[ht]
\centering
\caption{Mean bias and root mean squared error for structural model parameters under Design B across 100 replications ($J = 25$).}
\label{tab:struct-params-allN-B}
\begin{tabular}{llcccccc}
\toprule
& & \multicolumn{2}{c}{$\pi = 0.10$}
& \multicolumn{2}{c}{$\pi = 0.30$}
& \multicolumn{2}{c}{$\pi = 0.50$} \\
\cmidrule(lr){3-4}\cmidrule(lr){5-6}\cmidrule(lr){7-8}
$N$ & Parameter & Bias & RMSE & Bias & RMSE & Bias & RMSE \\
\midrule
\multirow{3}{*}{500}
& $\pi$      & $-$0.007 & 0.059 & $-$0.006 & 0.076 & $-$0.002 & 0.089 \\
& $\mu_1$    & 0.005 & 0.104 & $-$0.005 & 0.105 & $-$0.003 & 0.088 \\
& $\sigma_1$ & 0.020 & 0.101 & 0.004 & 0.103 & 0.015 & 0.090 \\
\midrule
\multirow{3}{*}{1000}
& $\pi$      & 0.003 & 0.053 & $-$0.002 & 0.060 & $-$0.002 & 0.066 \\
& $\mu_1$    & $-$0.001 & 0.106 & 0.007 & 0.097 & $-$0.005 & 0.081 \\
& $\sigma_1$ & $-$0.006 & 0.094 & 0.009 & 0.097 & 0.012 & 0.084 \\
\midrule
\multirow{3}{*}{3000}
& $\pi$      & $-$0.009 & 0.046 & $-$0.008 & 0.050 & $-$0.009 & 0.070 \\
& $\mu_1$    & $-$0.011 & 0.109 & $-$0.004 & 0.092 & $-$0.004 & 0.086 \\
& $\sigma_1$ & $-$0.011 & 0.097 & 0.014 & 0.110 & $-$0.008 & 0.083 \\
\bottomrule
\end{tabular}
\end{table}

Table \ref{tab:classif-allN-B} reports respondent classification accuracy and false positive rates for DIF detection under Design B. Since no DIF items are present, TPR is not applicable. The respondent classification error increases with $\pi$, from about 0.10 when $\pi = 0.10$ to about 0.37 when $\pi = 0.50$, with little variation across sample sizes. The AUC is stable across conditions, remaining close to 0.67. The lower AUC values compared to Design A reflect the absence of DIF, which in Design A provides additional separation between classes. The FPR is small across scenarios, indicating very limited spurious DIF detection.

\begin{table}[ht]
\centering
\caption{Respondent classification accuracy and false positive rates under Design B across 100 replications ($J = 25$).}
\label{tab:classif-allN-B}
\begin{tabular}{llccc}
\toprule
$N$ & Metric & $\pi = 0.10$ & $\pi = 0.30$ & $\pi = 0.50$ \\
\midrule
\multirow{3}{*}{500}
& Classification error & 0.102 & 0.317 & 0.376 \\
& AUC                  & 0.672 & 0.675 & 0.679 \\
& FPR                  & 0.002 & 0.016 & 0.009 \\
\midrule
\multirow{3}{*}{1000}
& Classification error & 0.101 & 0.308 & 0.371 \\
& AUC                  & 0.679 & 0.672 & 0.675 \\
& FPR                  & 0.021 & 0.026 & 0.012 \\
\midrule
\multirow{3}{*}{3000}
& Classification error & 0.101 & 0.304 & 0.373 \\
& AUC                  & 0.670 & 0.674 & 0.677 \\
& FPR                  & 0.112 & 0.046 & 0.029 \\
\bottomrule
\end{tabular}
\end{table}

\section{Empirical Analysis}\label{sec:empirical}

\subsection{First Empirical Illustration}

\subsubsection{Data Description}

We begin illustrating the proposed framework using data from the Trends in International Mathematics and Science Study (TIMSS). The analysis is based on Booklet 3 of dichotomously scored mathematics items administered to students from Singapore and Morocco. Singapore and Morocco represent countries at substantially different ends of the TIMSS mathematics proficiency distribution, a condition that \cite{huang2024investigating} identify as most susceptible to asymmetry-induced DIF under symmetric IRT models, which makes this dataset a substantively motivated testbed for the proposed framework. After aligning item sets across forms, the final dataset consists of $N = 605$ respondents and $J = 25$ items.

\subsubsection{Model Specification and Estimation}

We fit CLL models with one, two, and three latent classes, where the one-class model serves as a baseline without any latent heterogeneity. The multi-class models allow for class-specific Gumbel distributions through $(\mu_k, \sigma_k)$ and class-specific DIF parameters $\delta_{jk}$. The CLL link is adopted on both substantive and empirical grounds. Substantively, the 25 items in this dataset are mathematics items from Grade 4, predominantly multiple-choice in format. Multiple-choice mathematics items are susceptible to disjunctive response processes: a correct response may be achieved through multiple pathways including genuine mastery, strategic elimination of distractors, or proficiency-based guessing. Such a response process is argued to be associated with negative ICC asymmetry \citep{samejima2000logistic, lee2018alternative, huang2024investigating}, which the CLL link is designed to capture \citep{shim2023a}. The estimation uses the two-stage procedure in which $\lambda$ is selected by BIC over a grid of candidate values, followed by a confirmatory refit of the selected model. Table~\ref{tab:timss_bic} reports the BIC values for the fitted models. The two-class model is selected by the BIC. The three-class model yields a higher BIC despite additional flexibility. Inspection of the estimates shows that the third class has a very small mixing proportion, indicating that it does not represent a substantively distinct group.

\begin{table}[ht]
\centering
\begin{tabular}{lc}
\toprule
Model & BIC \\
\midrule
One-class & 15677.41 \\
Two-class & 15637.43 \\
Three-class & 15653.26 \\
\bottomrule
\end{tabular}
\caption{BIC values for the TIMSS data.}
\label{tab:timss_bic}
\end{table}

\subsubsection{Two-Class Model Results}

Table~\ref{tab:timss_2class_params} reports the parameter estimates for the selected two-class model.

\begin{table}[ht]
\centering
\small
\begin{tabular}{lrr}
\toprule
Item & $\hat{d_j}$ & $\hat{\delta}_{j1}$ \\
\midrule
M041010 & -0.178 & 0.000 \\
M041098 &  0.501 & 0.000 \\
M041064 &  0.122 & 0.000 \\
M041003 &  0.158 & 0.000 \\
M041104 &  0.461 & 0.000 \\
M041299 & -0.050 & 0.000 \\
M041329 & -0.420 & 0.000 \\
M041143 & -0.369 & 0.000 \\
M041158 &  0.065 & 0.000 \\
M041328 & -0.485 & 0.000 \\
M041155 &  0.337 & 0.124 \\
M041335 & -0.533 & 0.000 \\
M041184 & -0.392 & 0.000 \\
M051205 & -0.528 & 0.000 \\
M051039 &  0.010 & 0.000 \\
M051055 &  0.350 & 0.000 \\
M051006 &  0.359 & 0.000 \\
M051070 & -0.264 & 0.000 \\
M051018 &  0.532 & 0.000 \\
M051407 & -0.044 & 0.000 \\
M051410 &  0.370 & 0.000 \\
M051059 & -0.533 & 0.000 \\
M051093 &  0.027 & 0.000 \\
M051134 &  0.265 & 0.000 \\
M051077 &  0.182 & 0.000 \\
\bottomrule
\end{tabular}
\caption{Parameter estimates for the two-class model.}
\label{tab:timss_2class_params}
\end{table}

The structural parameter estimates are
\[
\hat{\nu}_1 = 0.735, \quad
\hat{\mu}_1 = -0.5, \quad
\hat{\sigma}_1 = 1.
\]

The empirical results thus support a two-class model. However, the estimated class structure is primarily characterised by differences in the latent distribution parameters $(\mu_k, \sigma_k)$. Only one item, M041155, has a non-zero DIF estimate and all remaining items are treated as anchors. This pattern (i.e., substantial latent impact with minimal item-level DIF) shows certain consistency with the findings of \cite{huang2024investigating}, which demonstrated that a meaningful proportion of cross-national DIF observed in TIMSS Mathematics assessments under symmetric IRT models can be attributed to ICC asymmetry interacting with between-country proficiency separation rather than genuine item bias. By specifying the CLL link within the mixture model, the present analysis likely accommodates this asymmetric component at the model level, which may resolve the portion of apparent DIF that would otherwise emerge under a symmetric specification, leaving only genuine item-level effects, of which there appears to be very little in this dataset.

\begin{figure}[ht]
\centering
\includegraphics[width=0.8\textwidth]{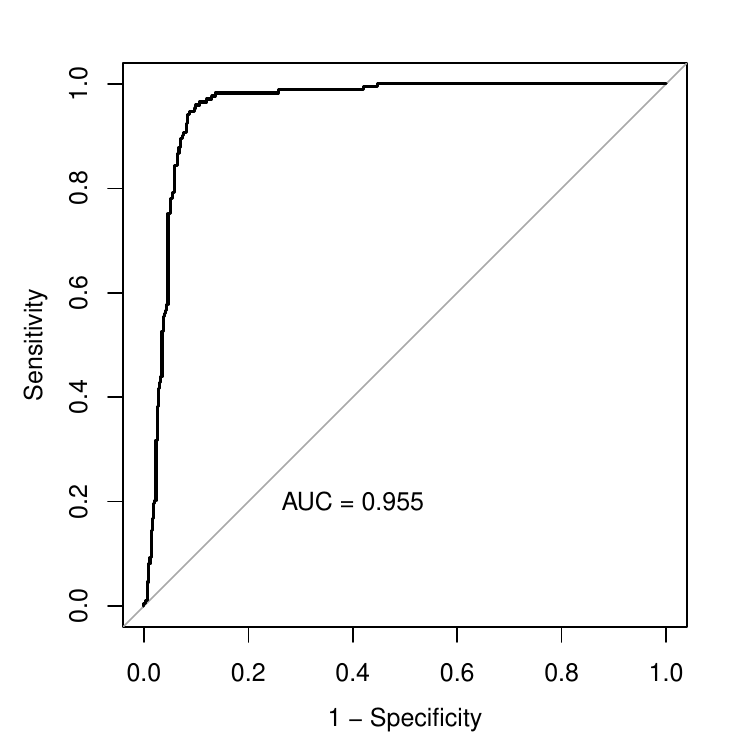}
\caption{ROC curve for classifying country membership using posterior class probabilities from the two-class model.}
\label{fig:roc_timss}
\end{figure}

To quantify the alignment between the latent classes and observed country, we compute the area under the ROC curve using the posterior class probabilities as a classifier. See Figure \ref{fig:roc_timss}. The resulting value is $\mathrm{AUC} = 0.955$, indicating that the latent class structure aligns closely with the observed country grouping.

\begin{figure}[ht]
\centering
\includegraphics[width=0.8\textwidth]{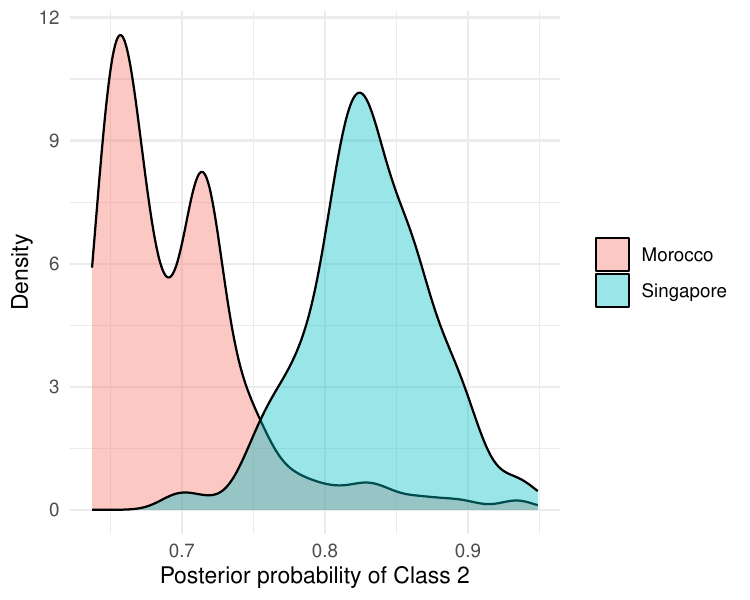}
\caption{Density of posterior probabilities of belonging to the focal class, by country.}
\label{fig:posterior_density_timss}
\end{figure}

The distribution of posterior class probabilities by country is shown in Figure~\ref{fig:posterior_density_timss}. The two distributions are clearly separated, with Singapore exhibiting higher posterior probabilities on average than Morocco. At the same time, there is some overlap between the distributions, indicating that the latent class structure captures clear but not complete separation between the two populations. The posterior probabilities are concentrated away from 0.5, suggesting that class membership is estimated with relatively high certainty for most individuals. This indicates that the latent classes are well separated in terms of likelihood, although the overlap between countries shows that this separation is not complete.

\subsection{Second Empirical Illustration}

\subsubsection{Data Description}
The second empirical illustration is based on a mathematics test from a midwestern university in the United States. This data set has been used in previous DIF studies, including \cite{bolt2002item}, \cite{de2011explanatory}, and \cite{wallin2024dif}.
The dataset consists of $N = 3000$ respondents and $J = 26$ dichotomously scored items. As in the previous empirical example, group membership is not observed and is treated as latent. The aim of this analysis is to examine whether the proposed framework can identify both latent heterogeneity and item-level DIF in a setting where such effects are expected to be present.

\subsubsection{Model Specification and Selection}

We fit one-, two-, and three-class versions of the proposed CLL-Gumbel model. Model comparison is based on the BIC computed from the confirmatory refit, see Table \ref{tab:emp2_bic}. The two-class model provides a better fit than both the one- and three-class model.

\begin{table}[ht]
\centering
\begin{tabular}{lc}
\toprule
Model & BIC \\
\midrule
One-class & 124187.5 \\
Two-class & 122623.3 \\
Three-class & 123522.4  \\
\bottomrule
\end{tabular}
\caption{BIC values for the second empirical dataset.}
\label{tab:emp2_bic}
\end{table}

\subsubsection{Two-Class Model Results}

The best-fitting two-class model yields the following structural parameter estimates:
\[
\hat{\pi}_1 = 0.770, \quad
\hat{\mu}_1 = -1.5, \quad
\hat{\sigma}_1 = 0.7.
\]

The item-level parameter estimates are reported in Table~\ref{tab:emp2_item_params}. The DIF estimates indicate several items with non-zero effects, including both positive and negative shifts. Specifically, items 7, 20--26 exhibit negative DIF effects, while items 11, 14, and 17 exhibit positive DIF effects. These results suggest the presence of both latent impact and item-level DIF, in contrast to the first empirical example where DIF effects were largely absent.

\begin{table}[ht]
\centering
\small
\begin{tabular}{cccccc}
\toprule
Item & $d_j$ & $\delta_j$ & Item & $d_j$ & $\delta_j$ \\
\midrule
1  & -0.325 &  0.000 & 14 & -0.300 &  0.117 \\
2  &  0.647 &  0.000 & 15 &  0.088 &  0.000 \\
3  &  0.427 &  0.000 & 16 & -0.492 &  0.000 \\
4  & -0.319 &  0.000 & 17 &  1.258 &  0.124 \\
5  &  1.118 &  0.000 & 18 & -0.199 &  0.000 \\
6  &  1.221 &  0.000 & 19 &  1.272 &  0.000 \\
7  & -0.403 & -0.155 & 20 &  0.357 & -0.179 \\
8  &  0.961 &  0.000 & 21 &  1.281 & -0.199 \\
9  &  0.214 &  0.000 & 22 & -0.625 & -0.149 \\
10 &  0.412 &  0.000 & 23 & -0.078 & -0.207 \\
11 & -0.612 &  0.129 & 24 & -0.520 & -0.215 \\
12 & -0.166 &  0.000 & 25 &  0.335 & -0.250 \\
13 &  0.743 &  0.000 & 26 & -0.080 & -0.239 \\
\bottomrule
\end{tabular}
\caption{Item parameter estimates for the two-class CLL-Gumbel model. The parameter $d_j$ denotes item difficulty in the reference class, and $\delta_j$ denotes the estimated DIF effect for the focal class.}
\label{tab:emp2_item_params}
\end{table}

To assess how the findings compare with a symmetric IRT specification, we also contrast the results with those reported in \cite{wallin2024dif} that were based on a two-parameter logistic (2-PL) model. The 2-PL analysis identified DIF for items 20--26, with all non-zero effects being negative. The CLL-Gumbel model also identifies items 20--26 as DIF items, and the estimated effects are again negative. Thus, the DIF pattern is stable across the symmetric and asymmetric specifications. The CLL-Gumbel model additionally identifies items 7, 11, 14, and 17 as DIF items. Items 11, 14, and 17 have positive DIF effects, whereas item 7 has a negative effect. The overlap on items 20--26 however suggests that the main DIF signal is not an artefact of the asymmetric specification. This pattern suggests that the asymmetric specification is sensitive to item-level effects that the symmetric model absorbs into its latent metric rather than attributing to differential functioning, consistent with the theoretical account of \cite{bolt2022item} and the empirical findings of \cite{huang2024investigating}. The overlap on items 20-26 across both specifications provides convergent validity for the DIF detections, while the additional items identified under the CLL-Gumbel model represent the incremental contribution of the asymmetric specification.

\section{Discussion}\label{sec:discussion}

We have proposed a general framework for DIF analysis under identifiable asymmetric IRT models that simultaneously addresses three challenges for large-scale assessment and survey measurement: unknown comparison groups, unknown anchor items, and asymmetric item response behaviour. We develop the framework concretely for the CLL link by embedding the CLL-Gumbel IRT model within a latent-class mixture and imposing $\ell_1$ regularisation on the class-specific difficulty shifts, achieving simultaneous identification of latent classes and DIF items without requiring pre-specified anchors or observed group labels. The same construction applies to other asymmetric IRT models that are themselves identifiable, such as those based on the negative log-log link \citep{shim2024parsimonious}.

A concern about any latent-class DIF method is the difficulty of distinguishing measurement non-invariance from distributional heterogeneity. A focal class with non-zero $\hat\delta_{jk}$ may represent a genuine subgroup for whom certain items are biased, or it may simply be a way for the model to improve fit by partitioning a continuous distribution into two discrete groups, with the DIF parameters absorbing the resulting item-level discrepancies. This ambiguity is irreducible without external information about class membership, and is not resolved by the sparsity assumption alone. What the sparsity assumption does is constrain the DIF attribution to a small number of items, ensuring that the scale is anchored and that the focal class is not defined entirely by its item-level differences from the reference class. This is a necessary but not sufficient condition for a substantive interpretation of the DIF items. 

The practical value of latent-class DIF analysis therefore depends critically on whether the recovered classess correspond to interpretable subgroups. In the TIMSS application, the two comparison countries are observed, allowing the latent class structure to be partially validated against know country labels by computinng the AUC of the posterior class probabilities used as a classifier of country membership. In setting where no external validation label is available (e.g., latent classes intended to capture speededness, aberrant behaviors, or response style shifts), practitioners must rely on other diagnostics, such as the substantive interpretability of the DIF pattern itself. We encourage users of the framework to treat latent-class DIF findings as exploratory rather than confirmatory in the absence of such external validation. At the same time, by explicitly accommodating ICC asymmetry within the measurement model, the framework reduces the risk that asymmetry-induced item differences are misattributed to differential functioning. DIF effects retained after applying this accommodation are therefore more likely to reflect genuine group-level item bias, providing a substantively cleaner starting point for follow-up content review and diagnostic investigation.

We acknowledge that several limitations of the current framework, each of which points to a natural direction for future research. The current framework does not provide confidence intervals or hypothesis tests for individual DIF effects. The procedure of \citet{chen2023dif},
which approximates the sampling distribution of $\hat\delta_{jk}$ by Monte Carlo simulation under the selected model when comparison groups are observed, applies in principle to the present case. The simulation study evaluates the proposed framework under the assumption that the data-generating model is correctly specified, that is, the true latent distribution is Gumbel and the true item response function follows the CLL link. Three misspecification conditions are of particular interest for future investigation. First, the latent distribution may be non-Gumbel: for instance, a normal mixture rather than a Gumbel mixture. This condition would quantify false positive rates in class and DIF recovery under distributional misspecification, and it is plausible in many applied settings where there is no substantive reason to expect Gumbel-distributed latent traits. Second, the true ICC asymmetry may be positive rather than negative. For example, data generated from a negative log-log model fitted with the CLL-Gumbel model. This would assess robustness to asymmetry direction misspecification, which is relevant because practitioners have to choose a link function before seeing results. Third, data generated from the CLL-Gumbel model but fitted with a symmetric 2-PL would quantify the cost of ignoring ICC asymmetry in DIF detection. This has been studied in \cite{bolt2022item} and \cite{huang2024investigating} through a different asymmetric IRT model and with manifest groups, but whether the observations will extend to the latent group setting studied here remains an open question. 

Finally, we note that the asymmetric CLL score function has implications for which items drive anchor selection. As discussed in Section 2.1, the score function decays rapidly when an item is well below a respondent's ability level. In samples with predominantly high ability, such as the Singapore group in the TIMSS application, easy items contribute relatively little gradient signal and may therefore have limited influence on anchor selection. This suggests that when the sample ability distribution is concentrated well above the difficulty range of many items, the regularised estimator may rely disproportionately on items of moderate to high difficulty for scale identification. If those items also carry DIF, the anchor set may be both narrow and contaminated, with easy items too uninformative to compensate. Practitioners should therefore be attentive to the difficulty distribution of the identified anchor items, particularly in high-proficiency samples where the effective anchor set may be smaller than the number of non-flagged items suggests.

The results of this paper highlight that assumptions about the shape of the item response function play a central role in DIF detection when comparison groups are unobserved. By explicitly modelling this asymmetry within a latent-class framework, the proposed approach provides a more flexible basis for separating latent impact from item-level non-invariance in settings where both are present but not directly observable.

\pagebreak

\bibliography{bibliography}

@article{bolt2002item,
  author = {Bolt, Daniel M. and Cohen, Allan S. and Wollack, James A.},
  title = {Item Parameter Estimation Under Conditions of Test Speededness: Application of a Mixture Rasch Model with Ordinal Constraints},
  journal = {Journal of Educational Measurement},
  year = {2002},
  volume = {39},
  number = {4},
  pages = {331--348}
}

@article{feuerstahler2026defining,
  title={Defining asymmetry in item response theory},
  author={Feuerstahler, Leah M and Verkuilen, Jay and Setti, Fabio and Johnson, Peter},
  journal={British Journal of Mathematical and Statistical Psychology},
  year={2026},
  publisher={Wiley Online Library}
}

@article{gonzalez2025identifiability,
  title={Identifiability analysis of the fixed-effects one-parameter logistic positive exponent model},
  author={Gonz{\'a}lez, Jorge and Baz{\'a}n, Jorge and Curi, Mariana},
  journal={British Journal of Mathematical and Statistical Psychology},
  volume={78},
  number={2},
  pages={440--458},
  year={2025},
  publisher={Wiley Online Library}
}

@article{huang2024investigating,
  title={Investigating item complexity as a source of cross-national DIF in TIMSS math and science},
  author={Huang, Qi and Bolt, Daniel M and Lyu, Weicong},
  journal={Large-scale Assessments in Education},
  volume={12},
  number={1},
  pages={12},
  year={2024},
  publisher={Springer}
}

@article{bauer2020simplifying,
  title={Simplifying the assessment of measurement invariance over multiple background variables: Using regularized moderated nonlinear factor analysis to detect differential item functioning},
  author={Bauer, Daniel J and Belzak, William CM and Cole, Veronica T},
  journal={Structural Equation Modeling: a Multidisciplinary Journal},
  volume={27},
  number={1},
  pages={43--55},
  year={2020},
  publisher={Taylor \& Francis}
}

@article{belzak2020improving,
  title={Improving the assessment of measurement invariance: Using regularization to select anchor items and identify differential item functioning.},
  author={Belzak, William and Bauer, Daniel J},
  journal={Psychological Methods},
  volume={25},
  number={6},
  pages={673–690},
  year={2020},
  publisher={American Psychological Association}
}

@article{bock1981marginal,
  title={Marginal maximum likelihood estimation of item parameters: Application of an {EM} algorithm},
  author={Bock, R Darrell and Aitkin, Murray},
  journal={Psychometrika},
  volume={46},
  number={4},
  pages={443--459},
  year={1981},
  publisher={Springer}
}

@article{bazan2006skew,
  title={A Skew Item Response Model},
  author={Baz{\'a}n, Jorge L and Branco, M{\'a}rcia D and Bolfarine, Heleno},
  journal={Bayesian Analysis},
  volume={1},
  number={4},
  pages={861--892},
  year={2006}
}

@article{bolfarine2010bayesian,
  title={Bayesian estimation of the logistic positive exponent IRT model},
  author={Bolfarine, Heleno and Bazan, Jorge Luis},
  journal={Journal of Educational and Behavioral Statistics},
  volume={35},
  number={6},
  pages={693--713},
  year={2010},
  publisher={SAGE Publications Sage CA: Los Angeles, CA}
}

@article{chen2023dif,
  title={DIF statistical inference without knowing anchoring items},
  author={Chen, Yunxiao and Li, Chengcheng and Ouyang, Jing and Xu, Gongjun},
  journal={psychometrika},
  volume={88},
  number={4},
  pages={1097--1122},
  year={2023},
  publisher={Cambridge University Press \& Assessment}
}

@article{cho2016ncme,
  title={An {NCME} instructional module on latent {DIF} analysis using mixture item response models},
  author={Cho, Sun-Joo and Suh, Youngsuk and Lee, Woo-yeol},
  journal={Educational Measurement: Issues and Practice},
  volume={35},
  number={1},
  pages={48--61},
  year={2016},
  publisher={Wiley Online Library}
}

@article{de2011explanatory,
  title={Explanatory secondary dimension modeling of latent differential item functioning},
  author={De Boeck, Paul and Cho, Sun-Joo and Wilson, Mark},
  journal={Applied Psychological Measurement},
  volume={35},
  number={8},
  pages={583--603},
  year={2011},
  publisher={Sage Publications Sage CA: Los Angeles, CA}
}

@article{dempster1977maximum,
  title={Maximum likelihood from incomplete data via the {EM} algorithm},
  author={Dempster, Arthur P and Laird, Nan M and Rubin, Donald B},
  journal={Journal of the Royal Statistical Society: Series B (Methodological)},
  volume={39},
  number={1},
  pages={1--22},
  year={1977},
  publisher={Wiley Online Library}
}

@book{embretson2013item,
  title={Item Response Theory},
  author={Embretson, Susan E and Reise, Steven P},
  year={2013},
  publisher={Psychology Press}
}

@article{martin2006irt,
  title={IRT models for ability-based guessing},
  author={Mart{\'\i}n, Ernesto San and Del Pino, Guido and De Boeck, Paul},
  journal={Applied Psychological Measurement},
  volume={30},
  number={3},
  pages={183--203},
  year={2006},
  publisher={Sage Publications Sage CA: Thousand Oaks, CA}
}

@article{magis2015detection,
  title={Detection of differential item functioning using the lasso approach},
  author={Magis, David and Tuerlinckx, Francis and De Boeck, Paul},
  journal={Journal of Educational and Behavioral Statistics},
  volume={40},
  number={2},
  pages={111--135},
  year={2015},
  publisher={Sage Publications Sage CA: Los Angeles, CA}
}

@article{parikh2014proximal,
  title={Proximal algorithms},
  author={Parikh, Neal and Boyd, Stephen},
  journal={Foundations and Trends{\textregistered} in Optimization},
  volume={1},
  number={3},
  pages={127--239},
  year={2014},
  publisher={Now Publishers, Inc.}
}

@article{wang2021using,
  title={Using lasso and adaptive lasso to identify DIF in multidimensional 2PL models},
  author={Wang, Chun and Zhu, Ruoyi and Xu, Gongjun},
  journal={Multivariate Behavioral Research},
  volume={58},
  number={2},
  pages={387--407},
  year={2023},
  publisher={Taylor \& Francis}
}

@article{shao1997asymptotic,
  title={An asymptotic theory for linear model selection},
  author={Shao, Jun},
  journal={Statistica Sinica},
  pages={221--242},
  volume={7},
  number={2},
  year={1997},
  publisher={JSTOR}
}

@article{zhao2006model,
  title={On model selection consistency of {L}asso},
  author={Zhao, Peng and Yu, Bin},
  journal={The Journal of Machine Learning Research},
  volume={7},
  pages={2541--2563},
  year={2006},
  publisher={JMLR. org}
}

@article{tutz2015penalty,
  title={A penalty approach to differential item functioning in {R}asch models},
  author={Tutz, Gerhard and Schauberger, Gunther},
  journal={Psychometrika},
  volume={80},
  number={1},
  pages={21--43},
  year={2015},
  publisher={Springer}
}

@article{bolt2022item,
  title={Item complexity: A neglected psychometric feature of test items?},
  author={Bolt, Daniel M and Liao, Xiangyi},
  journal={Psychometrika},
  volume={87},
  number={4},
  pages={1195--1213},
  year={2022},
  publisher={Cambridge University Press \& Assessment}
}

@inproceedings{verkuilen2023gumbel,
  title={Gumbel-Reverse Gumbel (GRG) model: A new asymmetric IRT model for binary data},
  author={Verkuilen, Jay and Johnson, Peter J},
  booktitle={The Annual Meeting of the Psychometric Society},
  pages={165--175},
  year={2023},
  organization={Springer}
}

@article{shim2024parsimonious,
  title={Parsimonious item response theory modeling with the negative log-log link: The role of inflection point shift},
  author={Shim, Hyejin and Bonifay, Wes and Wiedermann, Wolfgang},
  journal={Behavior Research Methods},
  volume={56},
  number={5},
  pages={4385--4402},
  year={2024},
  publisher={Springer}
}

@article{samejima2000logistic,
  title={Logistic positive exponent family of models: Virtue of asymmetric item characteristic curves},
  author={Samejima, Fumiko},
  journal={Psychometrika},
  volume={65},
  number={3},
  pages={319--335},
  year={2000},
  publisher={Springer}
}

@article{wallin2024dif,
  title={DIF analysis with unknown groups and anchor items},
  author={Wallin, Gabriel and Chen, Yunxiao and Moustaki, Irini},
  journal={Psychometrika},
  volume={89},
  number={1},
  pages={267--295},
  year={2024},
  publisher={Cambridge University Press \& Assessment}
}

@article{lee2018alternative,
  title={An alternative to the 3PL: Using asymmetric item characteristic curves to address guessing effects},
  author={Lee, Sora and Bolt, Daniel M},
  journal={Journal of Educational Measurement},
  volume={55},
  number={1},
  pages={90--111},
  year={2018},
  publisher={Wiley Online Library}
}

@article{lee2018asymmetric,
  title={Asymmetric item characteristic curves and item complexity: Insights from simulation and real data analyses},
  author={Lee, Sora and Bolt, Daniel M},
  journal={Psychometrika},
  volume={83},
  number={2},
  pages={453--475},
  year={2018},
  publisher={Cambridge University Press \& Assessment}
}

@article{molenaar2015heteroscedastic,
  title={Heteroscedastic latent trait models for dichotomous data},
  author={Molenaar, Dylan},
  journal={Psychometrika},
  volume={80},
  number={3},
  pages={625--644},
  year={2015},
  publisher={Cambridge University Press \& Assessment}
}

@article{shim2023a,
  author  = {Shim, Hyejin and Bonifay, Wes and Wiedermann, Wolfgang},
  year    = {2023},
  title   = {Parsimonious Asymmetric Item Response Theory Modeling with the Complementary Log--Log Link},
  journal = {Behavior Research Methods},
  volume  = {55},
  number  = {1},
  pages   = {200--219},
  doi     = {10.3758/s13428-022-01824-5}
}

\pagebreak
\appendix

\section{Gradients for the Proximal Gradient Descent}\label{app:gradients}

We derive all gradients required by the M-step of the proximal EM algorithm. For
the CLL link the item response function is $\Pjk(\theta) = 1 - e^{-e^{z}}$ where
$z = \theta - d_j - \delta_{jk}$. The score factor is
\begin{equation}\label{eq:score-factor}
  s(\theta; d_j, \delta_{jk})
  = \frac{e^{z - e^z}}{\Pjk(\theta)\bigl(1-\Pjk(\theta)\bigr)}.
\end{equation}
Define the expected sufficient statistic residual at quadrature node $q$ in class
$k$ as $D_{jq}^{(k)} = O_{jq}^{(k)} - \Pjk(\rho_q) S_q^{(k)}$.

The gradient of the negative expected complete-data log-likelihood is as follows.
With respect to the item difficulty:
\begin{align}
  \frac{\partial(-Q)}{\partial d_j}
  &= -\frac{1}{N}\sum_{k=0}^{K}\sum_{q=1}^{G}
     D_{jq}^{(k)} \cdot s(\rho_q; d_j, \delta_{jk}). \label{eq:grad-d}
\end{align}
With respect to the DIF parameter for class $k \geq 1$:
\begin{align}
  \frac{\partial(-Q)}{\partial \delta_{jk}}
  &= -\frac{1}{N}\sum_{q=1}^{G}
     D_{jq}^{(k)} \cdot s(\rho_q; d_j, \delta_{jk}). \label{eq:grad-delta}
\end{align}
With respect to the focal class location:
\begin{align}
  \frac{\partial(-Q)}{\partial \mu_k}
  &= -\frac{1}{N}\sum_{j=1}^{J}\sum_{q=1}^{G}
     D_{jq}^{(k)} \cdot s(\rho_q; d_j, \delta_{jk}). \label{eq:grad-mu}
\end{align}
With respect to the focal class scale:
\begin{align}
  \frac{\partial(-Q)}{\partial \sigma_k}
  &= -\frac{1}{N}\sum_{j=1}^{J}\sum_{q=1}^{G}
     D_{jq}^{(k)} \cdot s(\rho_q; d_j, \delta_{jk}) \cdot \rho_q^{*}, \label{eq:grad-sigma}
\end{align}
where $\rho_q^{*} = (\rho_q - \mu_k)/\sigma_k$ is the standardised quadrature node.
Note that gradients~\eqref{eq:grad-d} and~\eqref{eq:grad-delta} have identical
functional form: the gradient with respect to $d_j$ sums over all classes, whereas
the gradient with respect to $\delta_{jk}$ sums only over class $k$. This symmetry
is a consequence of the additive difficulty parametrisation in~\eqref{eq:irf}.

\section{Line Search Procedure}\label{app:linesearch}

In each M-step inner iteration, the step size $\alpha$ is selected by backtracking
line search. Starting from $\alpha_0$ (carried forward from the previous outer
iteration and inflated by a factor of 1.15 to allow the step to grow), we halve
$\alpha$ up to $B_{\max} = 40$ times until the penalised objective satisfies the
Armijo condition $F^{(t+1)} \leq F^{(t)} + \epsilon_{\mathrm{allow}}(|F^{(t)}| + 1)$
with $\epsilon_{\mathrm{allow}} = 5 \times 10^{-3}$. If no step satisfies the
condition after 40 halvings, the current parameters are left unchanged for that inner
iteration. This conservative threshold prevents numerical instabilities at boundary
regions of the parameter space while preserving the descent property in the
well-conditioned interior.

\section{Soft-Thresholding and Proximal Operator}\label{app:softthresh}

The proximal operator of the scaled $\ell_1$ function
$G(\bs\delta) = \alpha\lambda\|\bs\delta\|_1$ evaluated at $\bs{u}$ is the
element-wise soft-thresholding function
\[
  \bigl[\prox_{\alpha\lambda G}(\bs{u})\bigr]_{jk}
  = \mathcal{S}_{\alpha\lambda}(u_{jk})
  = \sign(u_{jk})\max\!\bigl(|u_{jk}| - \alpha\lambda,\; 0\bigr).
\]
DIF parameters for items not in the candidate set $\mathcal{C}$ are pinned to zero
throughout and are excluded from the thresholding step. The $\delta_{jk}$ estimates
are additionally clipped to $[-3.0, 3.0]$ to prevent numerical overflow in the score
factor at extreme values of the CLL argument $z_{jk}$.

\section{Additional results for Design A}
\label{app:bias-rmse}

\begin{figure}[htbp]
\centering
\includegraphics[width=0.9\textwidth]{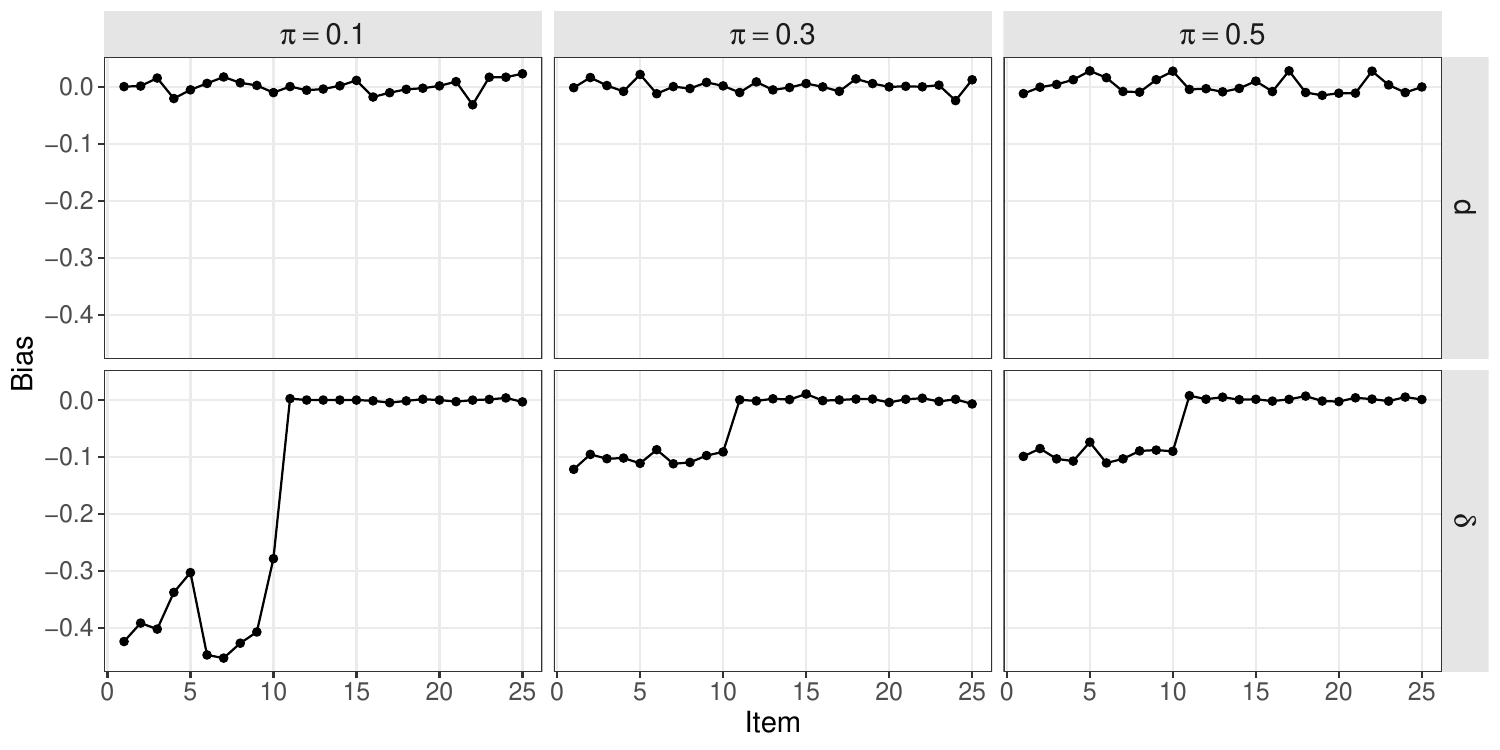}
\caption{Bias of the item parameters for $N=500$ across values of $\pi$ under Design A.}
\label{fig:bias_N500}
\end{figure}

\begin{figure}[htbp]
\centering
\includegraphics[width=0.9\textwidth]{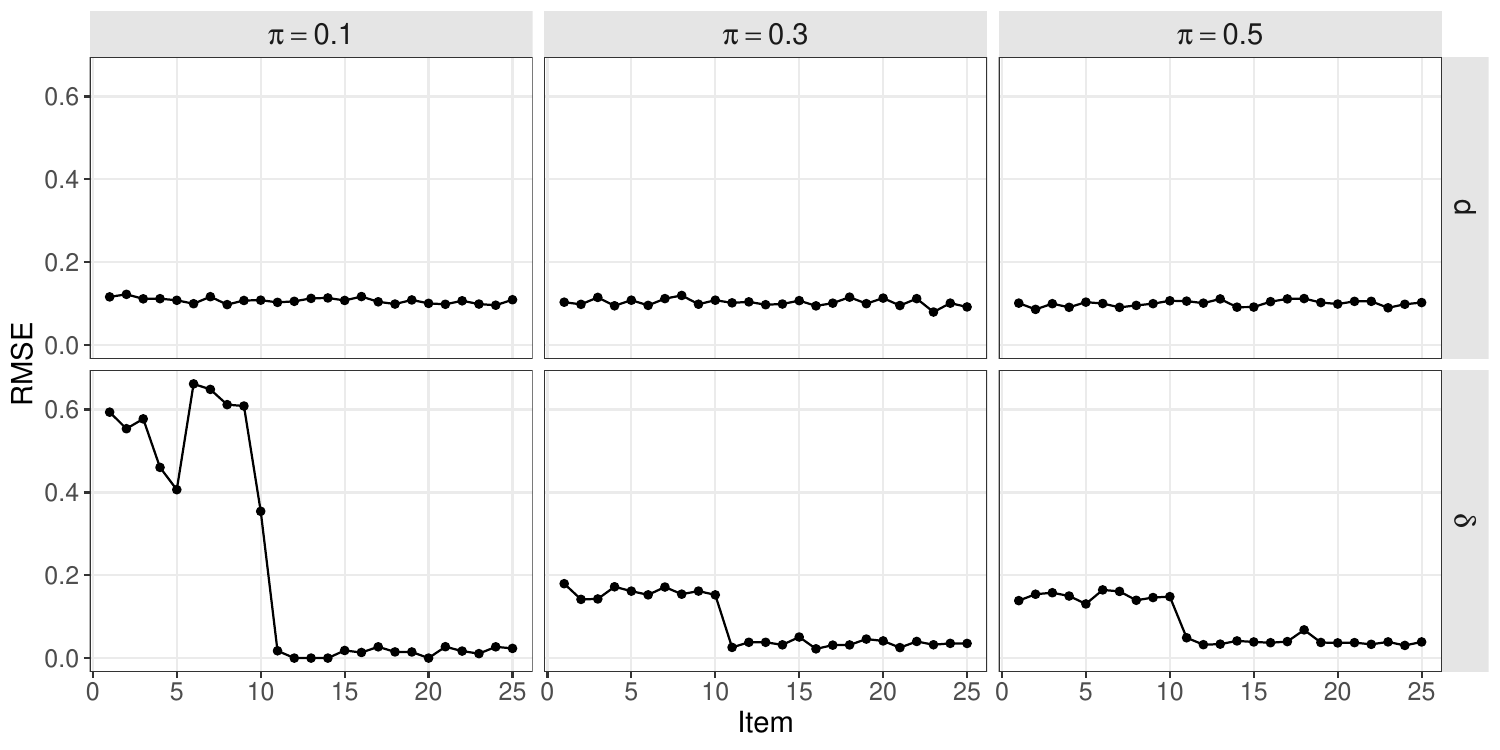}
\caption{RMSE of the item parameters for $N=500$ across values of $\pi$ under Design A.}
\label{fig:rmse_N500}
\end{figure}

\begin{figure}[htbp]
\centering
\includegraphics[width=0.9\textwidth]{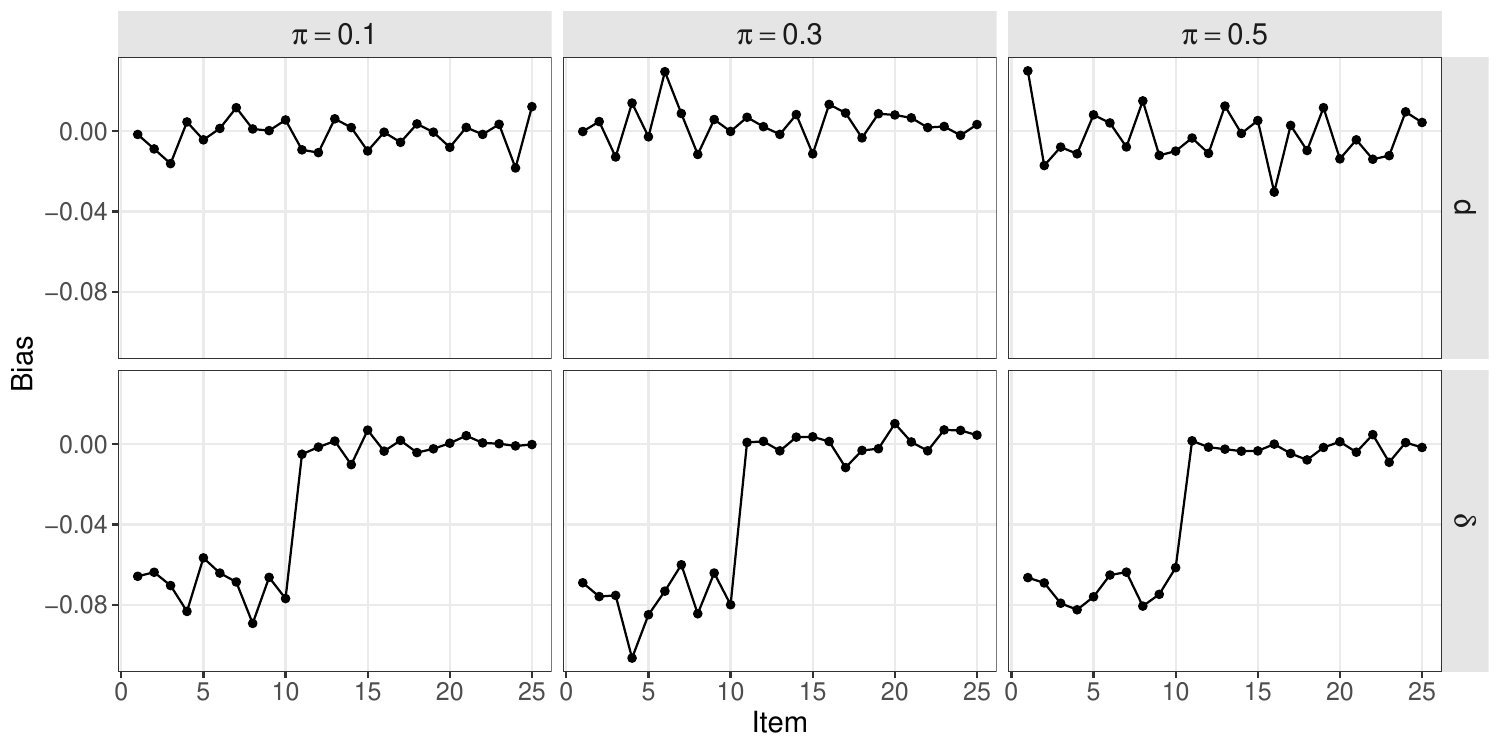}
\caption{Bias of the item parameters for $N=3000$ across values of $\pi$ under Design A.}
\label{fig:bias_N3000}
\end{figure}

\begin{figure}[htbp]
\centering
\includegraphics[width=0.9\textwidth]{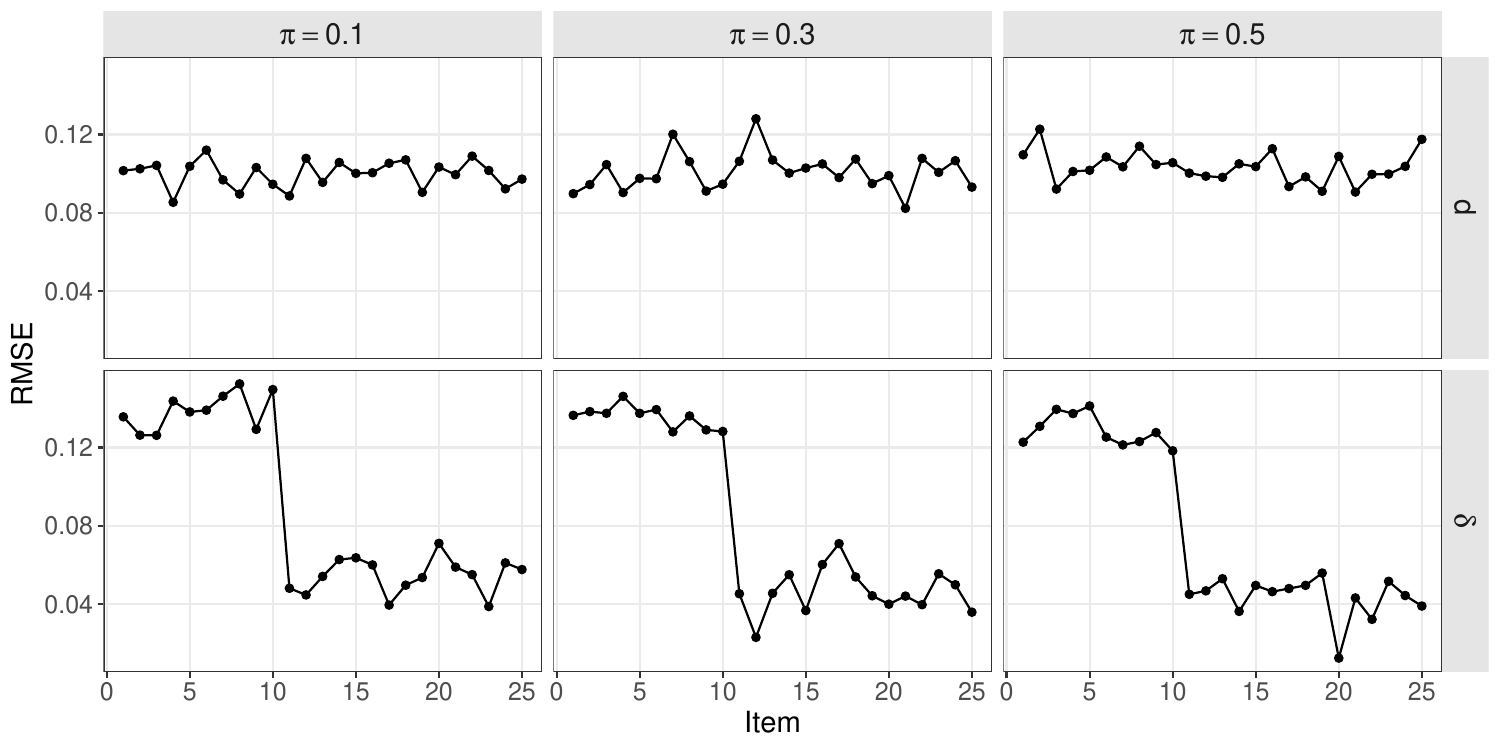}
\caption{RMSE of the item parameters for $N=3000$ across values of $\pi$ under Design A.}
\label{fig:rmse_N3000}
\end{figure}

\FloatBarrier

\section{Additional results for Design B}
\label{app:designB}

\begin{figure}[htbp]
\centering
\includegraphics[width=1\textwidth]{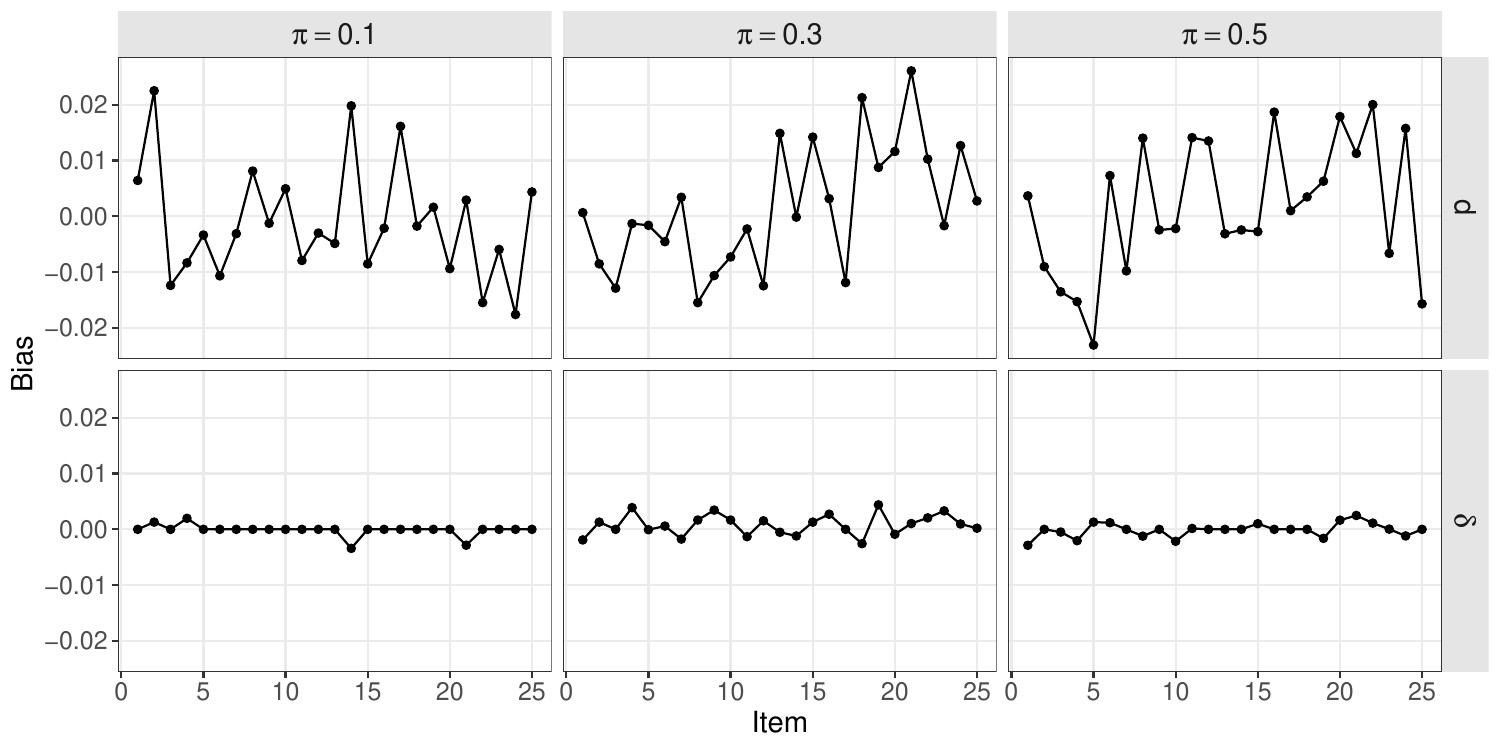}
\caption{Bias of the item parameters for $N=500$ under Design B across values of $\pi$.}
\label{fig:bias_N500_B}
\end{figure}

\begin{figure}[htbp]
\centering
\includegraphics[width=1\textwidth]{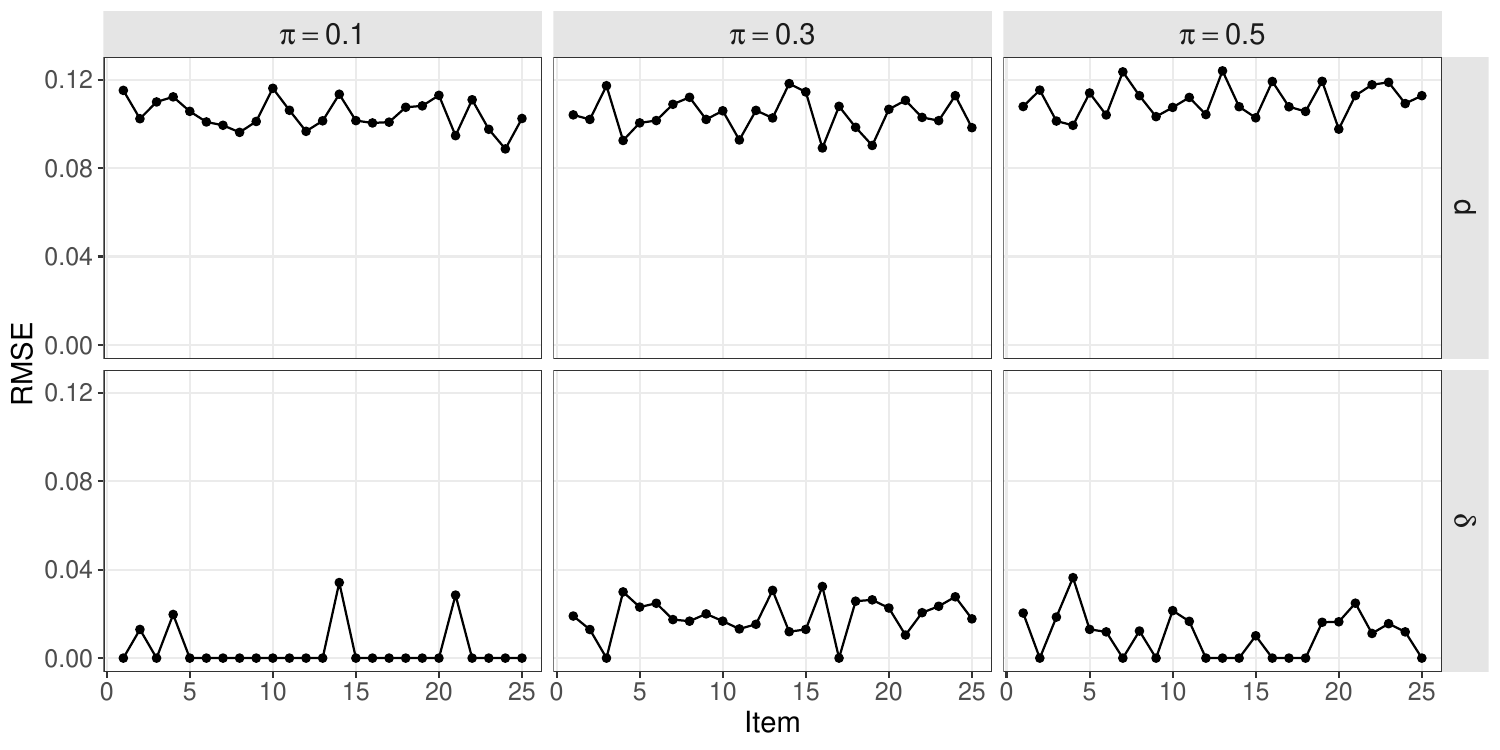}
\caption{RMSE of the item parameters for $N=500$ under Design B across values of $\pi$.}
\label{fig:rmse_N500_B}
\end{figure}

\FloatBarrier

\begin{figure}[htbp]
\centering
\includegraphics[width=1\textwidth]{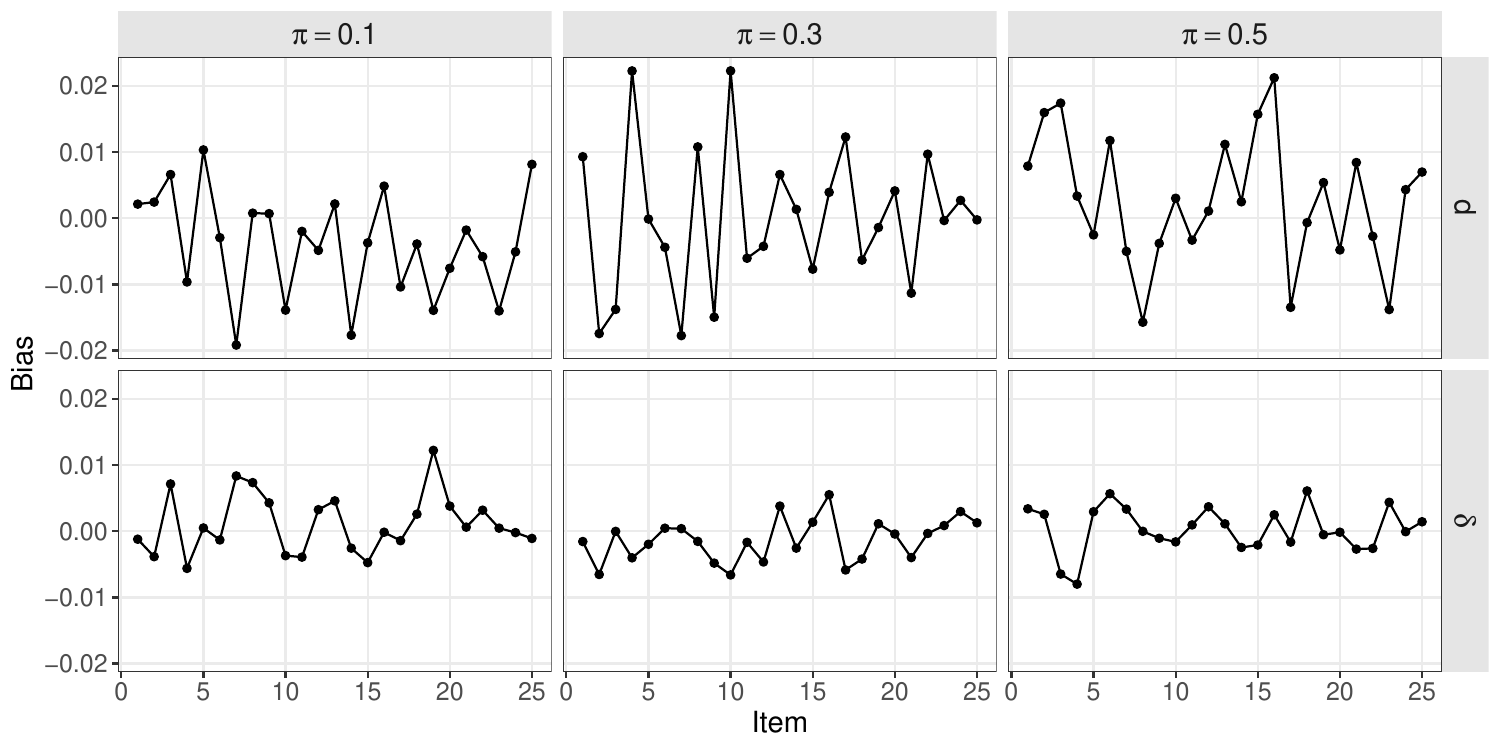}
\caption{Bias of the item parameters for $N=3000$ under Design B across values of $\pi$.}
\label{fig:bias_N3000_B}
\end{figure}

\begin{figure}[htbp]
\centering
\includegraphics[width=1\textwidth]{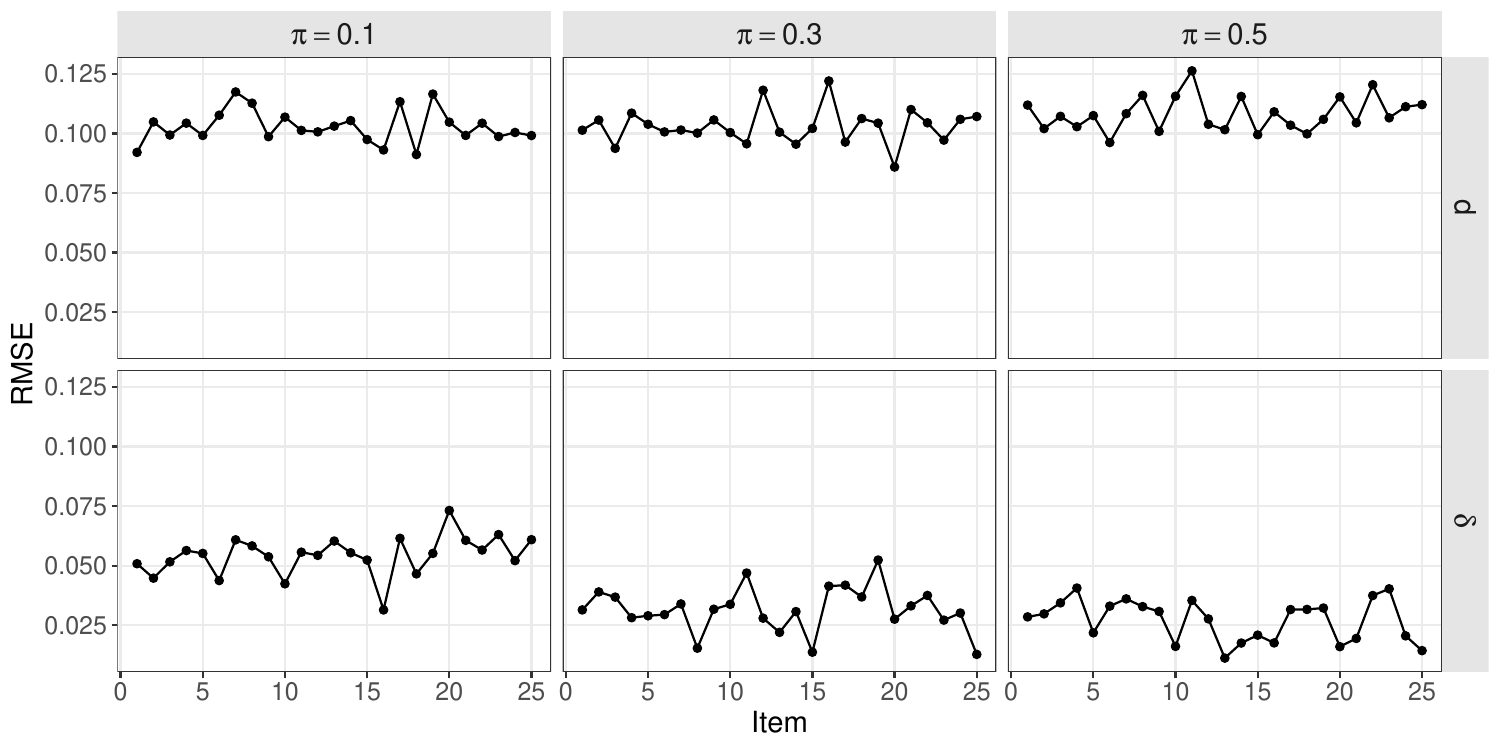}
\caption{RMSE of the item parameters for $N=3000$ under Design B across values of $\pi$.}
\label{fig:rmse_N3000_B}
\end{figure}

\end{document}